\documentclass[aps,prd,twocolumn,superscriptaddress,nofootinbib,floatfix,preprintnumbers]{revtex4-2}
\usepackage{graphicx}
\usepackage{amsmath,amssymb,mathtools,amsfonts}
\usepackage{bm}
\usepackage{siunitx}
\usepackage{hyperref}
\usepackage{physics}
\usepackage{color}
\usepackage{booktabs}
\usepackage{enumitem}

\newcommand{\EE}{\mathbb{E}}

\newcommand{\Bilby}{\texttt{Bilby}}
\newcommand{\GPU}{\texttt{GPU}}

\newcommand{\FFT}{\texttt{FFT}}

\newcommand{\thetabf}{\bm{\theta}}
\newcommand{\hbf}{\bm{h}}

\newcommand{\nbf}{\bm{n}}
\newcommand{\dbf}{\bm{d}}
\newcommand{\Cbf}{\bm{C}}
\newcommand{\Tbf}{\bm{T}}
\newcommand{\Rbf}{\bm{R}}
\newcommand{\Lbf}{\bm{L}}

\newcommand{\vct}[1]{\bm{#1}}        
\newcommand{\mat}[1]{\bm{#1}}        
\newcommand{\ddat}{\vct{d}}          
\newcommand{\hvec}{\vct{h}}          
\newcommand{\nvec}{\vct{n}}          
\newcommand{\cov}{\mat{C}}           
\newcommand{\cholL}{\mat{L}}         

\newcommand{\IUCAA}{\affiliation{Inter-University Centre for Astronomy and Astrophysics, Post Bag 4, Ganeshkhind, Pune 411 007, India}}
\newcommand{\ICTS}{\affiliation{International Centre for Theoretical Sciences - Tata Institute of Fundamental Research, Survey No. 151, Shivakote, Hesaraghatta Hobli, Bengaluru - 560 089, India.}}
\newcommand{\PSU}{\affiliation{Institute for Gravitation \& the Cosmos, Department of Physics, Penn State University, University Park PA 16802, USA}}

\begin{document}
\preprint{LIGO-P2600026}
\title{Accelerated Time-domain analysis for Gravitational Wave Astronomy}

\author{Vaishak Prasad}

\email{vaishakprasad@psu.edu}
\ICTS
\IUCAA
\PSU

\date{\today}

\begin{abstract}
Most current compact-binary searches and parameter-estimation pipelines evaluate the Gaussian-noise likelihood approximately using frequency-domain inner products with great success in analyzing gravitational wave signals. This is historically motivated by (i) the approximate stationarity of detector noise on sufficiently long timescales, allowing a circulant approximation in the domain that diagonalizes the noise covariance in the Fourier basis, and (ii) the efficiency of matched filtering via fast Fourier transforms. However, the advantage of frequency-domain analysis comes with its own limitations. In this article, we develop a self-contained, end-to-end, \emph{fully time-domain} formulation of gravitational-wave inference and present an implementation that makes the likelihood evaluation practical at scale by exploiting structured linear algebra, software, and hardware acceleration.
We validate the method using injections and demonstrate speedups for likelihood evaluation and on modern \GPU{}s.
We present \emph{tdanalysis}, an accelerated implementation that handles gaps, sharp boundaries, and multiple disjoint segments, and supports GPUs. We demonstrate some of its applications in gravitational wave astronomy.

\end{abstract}

\maketitle

\section{Introduction}
\label{sec:intro}
The LVK network of detectors has observed hundreds of binary black hole mergers and 2 confirmed neutron star mergers \cite{GWTC1, theligoscientificcollaboration2021gwtc21, GWTC3, LIGOScientific:2025hdt}. The gravitational wave signals recorded by the detectors are routinely used to infer astrophysical properties of the sources and, collectively, to discern deeper aspects of the universe, such as their distribution, formation channels, and cosmological properties (see e.g. \cite{LIGOScientific:2025pvj}). Gravitational wave observations are also used to test General Relativity itself \cite{Abbott2019_TGR_GWTC1, Abbott2021_TGR_GWTC2, Abbott2025_TGR_GWTC3, LIGOScientific:2025rid, LIGOScientific:2025obp}, which is an important undertaking that will help guide us in shedding light on important fundamental puzzles of the universe, like the validity no-hair conjecture, the laws of black hole mechanics and the interplay with quantum physics, dark matter and dark energy, the information loss paradox, and possible connections to quantum theory, etc. A crucial ingredient in this endeavor is the efficient search for gravitational wave signals in streaming observational data in gravitational wave detectors ~\cite{AdvancedLIGOCQG2015, AdvancedVirgoCQG2015, KAGRA2012}, the identification of potential candidates ~\cite{Usman2016PyCBC,Messick:2016aqy,Cannon2021GstLALSoftwareX,Babak2013ihope,Aubin2021MBTA,Chu2022SPIIR,Klimenko2016cWBMethod,Drago2021cWBSoftwareX,Robinet2020Omicron,Abbott2016GW150914Search}
and thereafter their data analysis and parameter estimation, i.e., the determination of the source properties.

The current framework for meeting these goals is the result of decades of research and development in the domain of signal processing ~\cite{Wiener1930GeneralizedHarmonic,Khintchine1934Korrelationstheorie,Schuster1898Periodogram,Bartlett1950PeriodogramContinuousSpectra,BlackmanTukey1958MeasurementPowerSpectra,CooleyTukey1965FFT,Welch1967SpectralEstimation,Whittle1953MultipleStationary}
and its adaptation to gravitational wave astronomy ~\cite{Sathyaprakash:1991mt, Dhurandhar:1992mw,LIGOScientific:2019hgc}, a Bayesian framework for the statistical interpretation ~\cite{Finn:1992wt, Finn:1992xs}, the adaptation and development of stochastic methods to sample high-dimensional posterior distributions ~\cite{Christensen:1998gf, ChristensenMeyer2001MCMCPE}, and also other physical aspects such as the success of numerical relativity ~\cite{Pretorius:2005gq, Campanelli:2005dd} and waveform modeling ~\cite{Cutler:1994ys, Buonanno:1998gg, Taracchini:2013rva, Ajith:2007qp, Field:2013cfa}. Today, the Frequency Domain (FD) approach is central to the data analysis strategy. 

The foundational detection and inference methods in gravitational-wave (GW) astronomy are built around a Gaussian-noise likelihood model \cite{Finn1992,Cutler:1994ys,JaranowskiReview}. In practice, compact-binary coalescence (CBC) searches and Bayesian parameter estimation have largely been performed in the \emph{frequency domain} \cite{Veitch2015LALInference,Usman2016PyCBCSearch,Ashton:2018jfp}. Assuming that (i) the noise in the detector is stationary, (ii) the domain is periodic, (iii) noise source distribution is Gaussian with mean zero, and that (iii) the noise adds linearly to the signals i.e. $d = s + n$, the noise covariance matrix is diagonal in the Fourier basis, and the set of diagonal components called the noise Power-Spectral Density, denoted by $S_n(f)$ completely characterizes the statistical properties of the noise. In practice, the likelihood is not exactly Gaussian, because the domain of analysis is not necessarily periodic. Given the assumption that the random process is wide-sense stationary, the periodicity exists in the weak sense: for long enough segments, the noise random process approximately repeats. Therefore, one adopts the Whittle approximation~\cite{Whittle1953MultipleStationary}, with the circulant assumption on the noise covariance matrix, and uses tapered windows to ensure that Fourier-domain methods can be applied stably.

The PSD is estimated in practice using neighboring data, employing Welch-style averaged periodograms \cite{Welch1967} or, more recently, a Bayesian fitting procedure involving splines and instrument line modeling ~\cite{Cornish2021BayesWavePipeline}.

In what follows, we use $T$ to denote the duration of the signal, $f_s$ the sampling rate, $N=f_s \times T$ to denote the number of time samples of the data vector being analyzed, the Big "O" notation to denote the computational complexity, and the "T" notation to denote the memory traffic.

The noise-weighted inner product in the frequency domain, called the matched filter, then takes a particularly simple form when the detector noise is approximately stationary and characterized by a one-sided power spectral density (PSD). The matched filter between the signal $s$ and template $h$ is defined as  \cite{Turin1960_MatchedFilters, Helstrom1968_StatisticalTheory, WainsteinZubakov1970_Extraction, Sathyaprakash:1991mt, Dhurandhar:1992mw}:
\begin{equation}
(s|h)\equiv 4\,\Re\int_{0}^{\infty}\frac{\tilde s(f)\,\tilde h^{*}(f)}{S_n(f)}\,df
\label{eq:innerprod}
\end{equation}
And the Signal to Noise Ratio (SNR) time-series is given by
\begin{equation}
    \rho(t)=\frac{(s|h_t)}{\sqrt{(h|h)}} \,,
    \qquad \tilde h_t(f)=\tilde h(f)\,e^{2\pi i f t}
\end{equation}
Given the noise PSD, the source properties of the gravitational wave signals are determined using matched filtering, comparing the signal against a precomputed bank of templates whose source properties are known ~\cite{Owen:1998dk}. Assuming General Relativity, several waveform models exist that are routinely used to predict gravitational wave signals from binary systems to varying degrees of accuracy.

The source parameters are then determined using a Bayesian approach. The source properties of binary systems, say for Binary Black Holes (BBH)s on quasi-circular orbits in GR, can be described using 15 parameters, which include 8 intrinsic parameters, viz. mass-ratio $q$, chirp-mass $\mathcal{M_c}$, and the two spin vectors $\Vec{S}_{1,2}$; 7 extrinsic parameters that determine the source's location and orientation in the sky with respect to the GW detectors: the right-ascension $r_a$, declination $\delta$ and luminosity distance $D_{L}$, polarization angle $\psi$, inclination $\theta_{jn}$ of the system wrt the line of sight, phase angle $\phi$ of the binary at reference time, and coalescence/ peak time $t_{c}$ that determines the arrival time of the signal. The waveform models used for matched filtering are thus labeled by these parameters $\thetabf$. In the Bayesian approach, one seeks to obtain a posterior distribution $p(\thetabf | d)$ for the values of these 15 source parameters, given prior knowledge and a Likelihood function. 

The Likelihood function for a candidate signal $\hbf(\thetabf)$ given the data $\dbf$ recorded in the detector is just the Gaussian likelihood function containing the inner product of the noise estimate $\nvec = \ddat-\hvec(\thetabf)$ with itself, and is conveniently computed in the frequency domain using the matched filter defined above:
\begin{equation}
\ln \mathcal{L}(d\,|\,\theta,H)=\text{const}-\frac{1}{2}\,\bigl(d-h(\theta)\, \big|\, d-h(\theta)\bigr)
\label{eq:td_like_def}
\end{equation}
Where $\hbf(\thetabf)$ is the template generated from a waveform model with parameters $\thetabf$.

Given the high-dimensional nature of the parameter space (15 for non-eccentric BBHs) and the complexity of the likelihood surface, it is practically infeasible to integrate the likelihood analytically or numerically to obtain the posteriors. To tackle this issue, stochastic sampling techniques like the Markov Chain Monte Carlo (MCMC) and its variants, nested sampling ~\cite{MetropolisEtAl1953,Hastings1970,Geyer1991ParallelTempering,SwendsenWang1986ReplicaMC,HukushimaNemoto1996ExchangeMC,MarinariParisi1992SimulatedTempering,Neal1996TemperedTransitions,Green1995RJMCMC,Skilling2004NestedSampling,Skilling2006NestedSamplingGeneral}, etc,  are used to efficiently obtain a representative sample from the posterior distribution, which can be used to conveniently estimate various integrals.

Each step in an iterative Monte Carlo procedure that aims to sample the posterior distribution involves repeatedly drawing a proposal sample from the prior distribution and evaluating the likelihood function defined above to determine acceptance. Historically, there are several advantages to adopting the frequency domain for gravitational wave data analysis. 
\begin{enumerate}
    \item given all the ingredients, the likelihood evaluation in the frequency domain amounts to a computational complexity (CC) of $O(n)$, assuming the waveforms are already in the Fourier basis, as opposed to a general covariance matrix with non-diagonal elements that amounts to $O(N^2)$.
    \item Due to the stationary phase approximation and phenomenological approaches, constructing waveform models in the frequency domain is inherently convenient and fast, which then can directly be used in the likelihood evaluations, without needing further transformations.
    \item The existence of Fast Fourier Transformation (FFT) algorithms allows one to transform a given time domain signal recorded in the detectors, and template waveforms if generated in the time domain, relatively inexpensively (compared to other operations) at $O(N \log N)$.
    \item Operating in the frequency domain also gives access to a mature battery of filters, windows, and tools already existing and implemented in frequency domain analysis.
    \item Due to the diagonal covariance of the noise, the memory traffic also scales as $T(N)$, and the memory requirements to store vectors instead of matrices are low. One only needs to store the PSD in memory in $T(N)$ space, along with an $N$-dimensional vector, which scales linearly with $N$.
\end{enumerate}
In computing, the adaptability and practicality of an algorithm are determined by its (i) computational complexity, (ii) memory requirement, and (iii) the memory bandwidth requirements. While (i) determines how long the CPU needs to finish processing the arithmetic instructions, (ii) and (iii) determine how fast the data required for executing those instructions can be streamed from storage to RAM, and from RAM to the CPU cache registers. In what follows, we analyze in detail the computational requirements for gravitational wave data analysis and discuss the resource budget, in an attempt to explain why FD methods gained popularity.
\begin{figure*}
    \centering
    \includegraphics[width=0.52\linewidth]{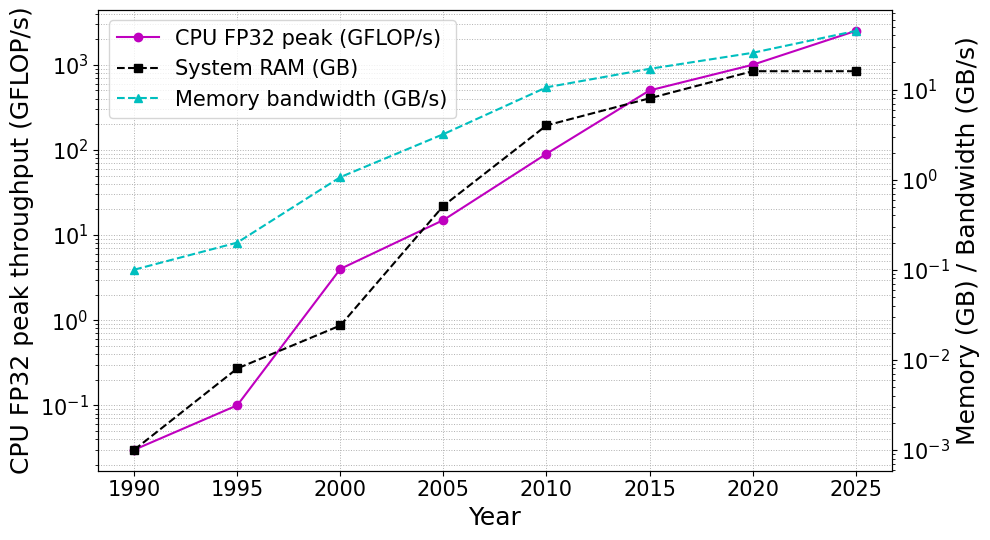}
    \includegraphics[width=0.47\linewidth]{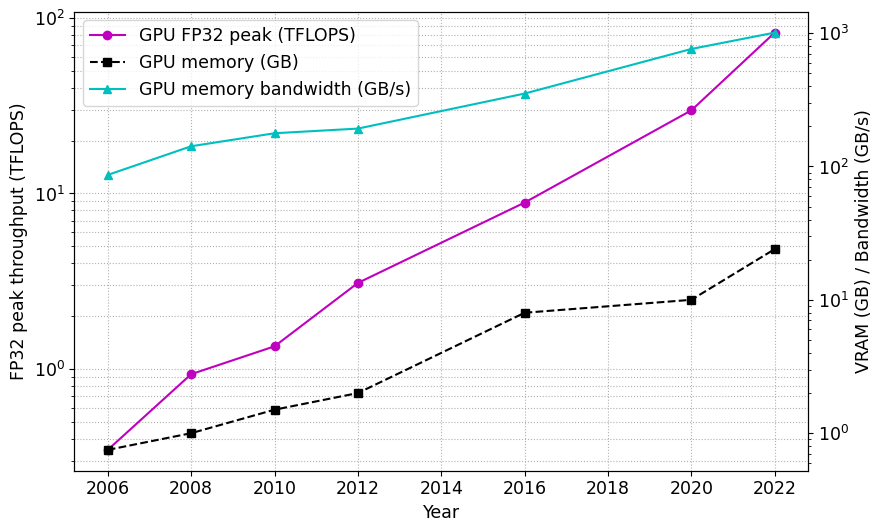}
    \caption{Average amount of CPU FLOPs for 32-bit arithmetic and the memory bandwidth since 1990}
    \label{fig:avgcomp}
\end{figure*}

Historically, when these methods were adapted to GW data analysis at the end of the 20th century, the best computers had a few to at most $O(100)$ Megabytes of Random Access Memory (RAM)~\ref{fig:avgcomp}. The memory bandwidth ranged from $T(100)$ MBps to $T(1)$ GBps. This meant that it was impractical to store dense matrices larger than $4096\times 4096$ in RAM, which costs $64$ MiB. Furthermore, streaming such a matrix from the RAM for computation would take O(1) time. This meant that it was infeasible to analyze signals longer than half a second, at a sampling rate of $4096 Hz$.

The frequency-domain approach was thus a widely accepted solution to this problem, garnering significant attention and success in its adaptation and, eventually, in the analysis of gravitational wave signals. The general strategy for signal processing and analysis was to formulate the problem at hand into the FD. The diagonal covariance matrix in the Fourier domain meant that it was no longer necessary to store and stream matrices, but just vectors, which require only $16KiB$ per vector for $N=16384$ with FP64. However, the FD approach also has some disadvantages. 

Operating in the FD and using the Fourier basis requires the data to be periodic and smooth. Any FD analysis thus begins by ensuring this, by applying windowing functions such as the Hann ~\cite{vonHann1903HandbookClimatology}, Hamming ~\cite{Hamming1977DigitalFilters}, and Tukey ~\cite{BlackmanTukey1958MeasurementPowerSpectra} windows, so that the waveforms are periodic, avoiding the problems associated with spectral leakage and making it difficult to treat signals localized in time. This severely limits the analysis of time-localized portions of waveforms that start abruptly and excludes certain portions from analysis, without compromising the ends of the segment. Furthermore, the windows attenuate the signal, causing a loss of SNR. On the other hand, the use of steep windows can gradually increase Gibbs oscillations and lead to spectral instabilities. For similar reasons, it is much more difficult to treat non-stationary, transient noise backgrounds in the frequency domain, since transforming to the Fourier domain requires a sufficiently large domain to capture the signal's low-frequency features, forcing one to assume that the statistical properties of the noise remain unchanged in the domain of support of the Fourier basis functions.

Despite the computational challenges, time-domain methods have been used in the past under various capacities and restrictions. The first pipeline to apply them to Gravitational Wave Astronomy was in detection infrastructure~\cite{Messick:2016aqy, Sachdev:2019}, enabling streaming low-latency matched-filter analysis. Since 2019, multiple pipelines and works exist that use a time-domain framework for carrying out black-hole spectroscopy and ringdown analysis~\cite{Carullo:2019flw, Isi:2019aib, Isi:2021iql, Isi:2023cmt, Carullo:2023rpl}. In ~\cite{Isi2021AreaLaw}, a time domain framework was adopted to analyze the inspiral and ringdown segments separately and test Hawking's area increase theorem~\cite{Hawking1971_AreaTheorem} in BBH mergers using GW150914. Later in ~\cite{Miller:2023ncs, Miller_2025}, individual segments of the time-domain signal are analyzed to extract physics using their \emph{tdinf} code~\cite{Miller:tdinf:2025}, keeping some of the extrinsic parameters fixed for speed. In ~\cite{LIGOScientific:2025rid}, a time-domain method was used to analyze individual segments of GW250114~\cite{LIGOScientific:2025rid, LIGOScientific:2025obp} and observationally establish the validity at unprecedented accuracy. Although there has been substantial progress in the adaptation of time-domain methods, their computational complexity is currently a roadblock in using them in the same capacity and runtime as frequency-domain methods. More recently, ~\cite{Sharma:2026} proposes heterodyning methods to accelerate time-domain analysis. They also independently discover one of the acceleration method implemented in this work.

In this work, we reanalyze the feasibility of adapting time-domain methods in the era of modern hardware, explore ways to accelerate them, and introduce an alternate implementation, called \emph{tdanalysis}, available for use upon request. We demonstrate the feasibility and performance of stochastic sampling of Bayesian posteriors using an explicit time-domain likelihood, and discuss how modern hardware infrastructure, such as vector multiprocessors,  \GPU{}s, and advancements in memory infrastructure, enable this practically when harnessed.

Our goal is \emph{not} to discard the stationary Gaussian likelihood, but to use an equivalent, systematically generalizable time-domain representation that (i) remains efficient and robust across realistic data-conditioning choices, (ii) allows one to analyze any localized portions of the signal in time domain, (iii) remove the need to apply window functions, (iv) realize the maximum speedups that one can achieve for prospectively analyzing long signals using GPUs. This will help set the stage for improved, feasible data analysis methods for future gravitational-wave detectors. We also demonstrate the validity by analyzing injected and real gravitational-wave events. 

We wish to point out that an older version of this pipeline was used in the Author's PhD thesis~\cite{MyThesis, Prasad:2023bwa}, where it was shown that the ringdown portion of the gravitational radiation infalling at the outer common dynamical horizon in BBH mergers encodes information about the source properties. This current version of the pipeline also finds use in the Multi-Segment Consistency Test (MSCT), presented recently~\cite{prasad_msct}, which proposes to analyze multiple time segments of the signal simultaneously to test for deviations from general relativity. Further, the pipeline presented here, along with the MSCT is also used to propose a more consistent way of measuring waveform systematics in any segment of the time-domain waveform~\cite{prasad_wfs}. These proposed applications, along with time-domain analysis of full gravitational wave signals, have only been possible due to the various acceleration methods developed here.

The plan of the paper is as follows. First, in Sec.~\ref{sec:likelihood}, we describe the statistical basis of the time-domain approach and the whitened representation. We also describe some of the salient features of the pipeline, including handling gaps and finite segments, which are seamless in the time-domain representation. In Sec.~\ref{sec:cost}, we carry out an algorithmic complexity analysis and discuss the feasibility of current and future gravitational wave signals in the context of modern computational hardware. In Sec.~\ref{sec:bench} we describe potential speedups, both algorithmic and hardware, that make time-domain methods feasible for gravitational wave astronomy. In Sec.~\ref{sec:results}, we demonstrate the time-domain pipeline using injections. We also validate and demonstrate its capabilities by applying it to the loudest GW event observed so far, GW250114. We summarize the results in Sec.~\ref{sec:disc}.

\section{Data model and time-domain likelihood}
\label{sec:likelihood}

Consider the signal incident on a GW detector. Assuming that the noise adds linearly to the signal, the strain data response of the detector can be expressed as
\begin{equation}
  d(t) = h(t;\bm{\theta}) + n(t), ,
\end{equation}
where $\hbf$ is the detector response to a CBC waveform parameterized by $\bm{\theta}$, and $\nbf$ the noise. Both can be represented in the form of row vectors of size $N$. Each time sample of these vectors will be denoted by a Latin index, e.g., $n_j$ for the $jth$ noise sample, with $j\leq N$.

To quantify the statistical properties of the noise $n(t)$, we will use expectation values $E(f(n))$, where the averages induced by $E$ correspond to the ensemble averages over different noise realizations. In practice, in the absence of access to multiple parallel universes, the assumption of ergodicity ~\cite{Birkhoff1931ErgodicTheorem} enables us to replace the ensemble averages with time-segment averages, especially since the noise in different time regimes of the detector, i.e., over large time scales, is expected to be uncorrelated. Thus, we assume that the noise is wide-sense stationary, as is commonly adopted in the frequency-domain approach. Time domain methods are generic and can be used to address a wide variety of noise source properties, and it is generally easier to relax these assumptions to handle non-stationarity under various approximations than in the frequency domain~\cite{Priestley1965EvolutionarySpectra,PriestleyRao1969TestNonstationarity,Dahlhaus1997FittingNonstationary,Dahlhaus2000LikelihoodApprox,GuinnessStein2013LocalStationaryGP,Kalman1960FilteringPrediction,VanBellegemVonSachs2008EWS}.

But for the purposes of this work, we continue to hold onto this assumption and discuss the feasibility of time-domain methods in the modern hardware scenario, leaving improvements to future work. Given the wide sense stationarity, the statistical properties of noise, and particularly the noise covariance $\Cbf$, depend only on the lag and not absolutely on the time index. $\cov_{ij}$ is then not completely dense and has a special structure:
\begin{equation}
    \cov_{ij} \equiv \EE[n_i n_j] = \rho\!\left(\lvert i - j \rvert\right)
\end{equation}
Where $\rho$ is defined as the autocorrelation function. Thus, $\cov_{ij}$ has a band structure and is symmetric. Such a matrix is called a Symmetric Toeplitz matrix \cite{GrenanderSzego1958,Gray2006Toeplitz, Khintchine1934Korrelationstheorie,BrockwellDavis1991TimeSeries}.

Apart from wide-sense stationarity, note that we have not yet imposed any constraints on the source distribution of $n(t)$.  The simplest of the choices, which is used in gravitational wave analysis, is that $n(t)$ is sourced by a multivariate Gaussian distribution, in which case the statistical properties of noise are completely characterized by its two non-zero cumulants: its mean $E[n(t)]$ (a trivial shift can be applied to set the mean to zero), and a noise covariance $E[n_i, n_j] = \cov_{ij}$ of size $N\times N$ where $N$ is the number of time samples. This greatly simplifies and motivates the inner product
\begin{equation}
    \langle s, h \rangle = \vct{s}^{\top}\,\cov^{-1}\,\vct{h}  \label{eqn:ip}
\end{equation}
The Spectral theorem in linear algebra ~\cite{Cauchy1829SecularEquation,HornJohnson2013MatrixAnalysis,Axler2015LinearAlgebraDoneRight} dictates that a real symmetric matrix admits an eigen-basis in which it is diagonal. If one further imposes periodic boundary conditions on the domain (and thus on the noise and signal), it can be shown that this basis is the Fourier Basis, and that $\cov_{ij}$ becomes circulant, leading to the familiar FD approach, where efficient FFT algorithms are available. In reality, the data and the random process are not periodic. One circumvents this by applying window functions, which taper the data to zero smoothly at both ends. This enables one to make the circulant assumption and reduce spectral leakage. The latter is important in making the inner product in Sec.~\ref{sec:likelihood} stable against small perturbations in the domain. However, the downside is that it attenuates the signal across time samples, as windows have non-local support in the time domain, making it difficult to analyze localized time segments with sharp cutoffs without special techniques \cite{Harris1978} or by compromising information at the boundaries.

In what follows, we relax the circulant assumption and remain in the time domain, thereby removing the need for tapering or windowing and enabling analysis of sharply terminated segments of the signal that start and/or end abruptly. 

Using the inner product defined in Eq.~\eqref{eq:innerprod}, one can define a Gaussian likelihood model:
\begin{equation}
\ln \mathcal{L}(d\,|\,\theta,H)=\text{const}-\frac{1}{2}\,\langle d-h(\theta),  d-h(\theta)\rangle \label{eqn:lik}
\end{equation}
Note that this is the exact Gaussian Likelihood, and no Whittle-like approximation is necessary, and hence inherently more precise than the frequency domain representation, especially for short signals.

\subsection{Whitened representation}
Analogous to the amplitude spectral density representation in the frequency domain, one can develop a whitened representation of the signals in the time domain by factoring the noise Covariance matrix $\Cbf$. A convenient choice is the Cholesky decomposition
\begin{equation}
    \cov = \cholL\cholL^\top
\end{equation}
The inner product in \eqref{eqn:ip} can then be written as
\begin{equation}
    \langle s, h \rangle = \bar{\vct{s}}^{\top}\,\bar{\vct{h}} 
\label{eq:td_innerprod_whitened}
\end{equation}
Where $\bar{v} = \cholL\cholL^{-1} v$ is the whitened representation of $v$ and $W = \cholL^{-1}$ can be called as the whitening operator. If $\nbf$ is a multivariate Gaussian, then it can be shown that $
\tilde{\nbf}$ is uncorrelated unit variance normally distributed, as long as $\mathbf{W}\Cbf\mathbf{W}^T = \mathbf{I}$. The whitened residuals are then $\bm{r} = \bm{W} (\ddat-\hvec)$, and Eqn.~\eqref{eqn:lik} takes the explicit form
\begin{equation}
  \ln \mathcal{L}(\bm{\theta})
  = -\frac{1}{2}
  \left[
    \bm{r}^{\top}\bm{r}
  \right] +C.
  \label{eq:td_like_whitened}
\end{equation}
%

\subsection{Implementation details}
\begin{figure}
    \centering
    \includegraphics[width=0.5\linewidth]{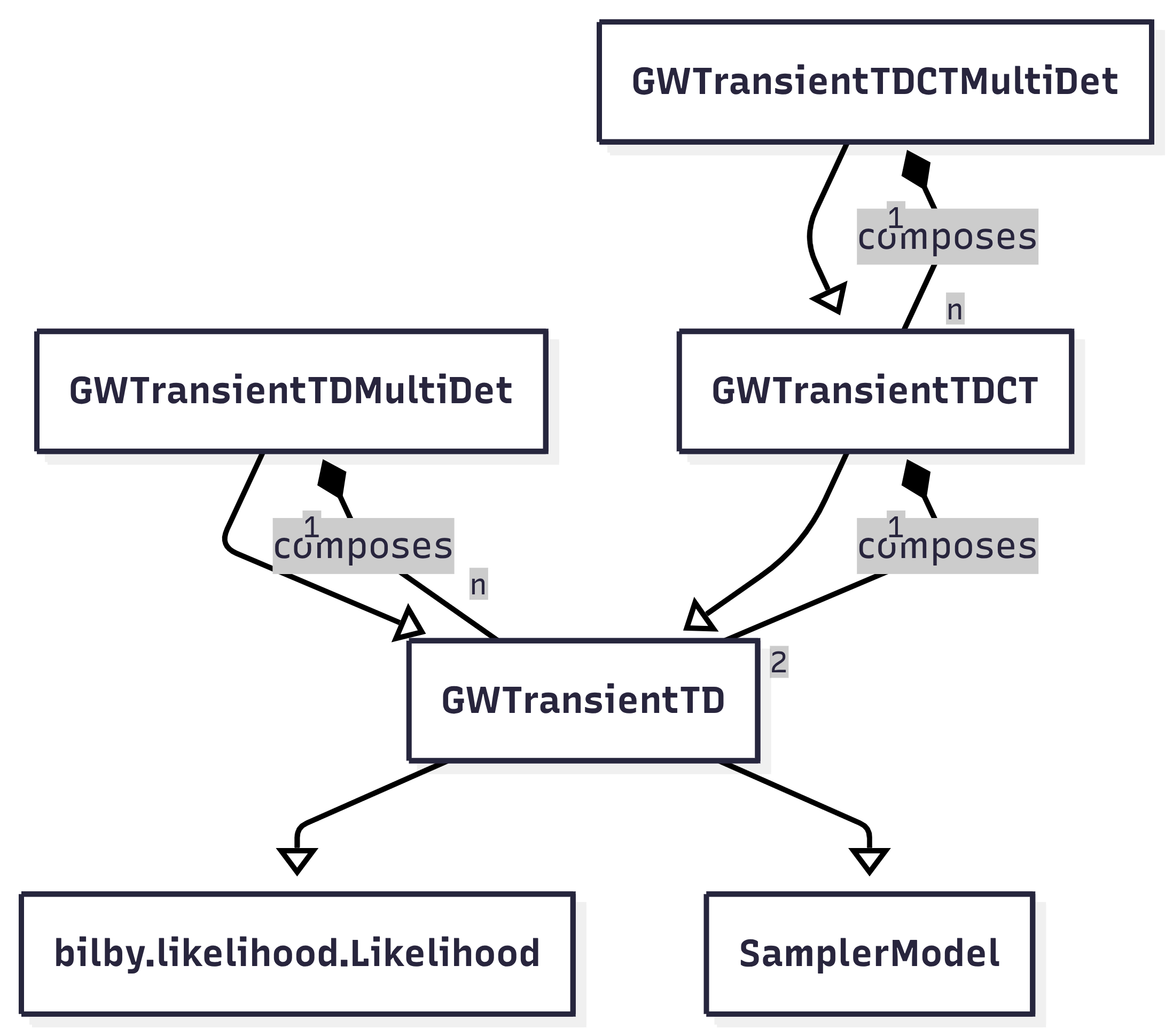}
    \caption{The \emph{tdanalysis} infrastructure for time domain analysis. This class diagram shows the basic skeleton of inheritance and composition for the major likelihood classes involved.}
    \label{fig:classd}
\end{figure}
\begin{figure*}
    \centering
     \includegraphics[width=0.49\linewidth]{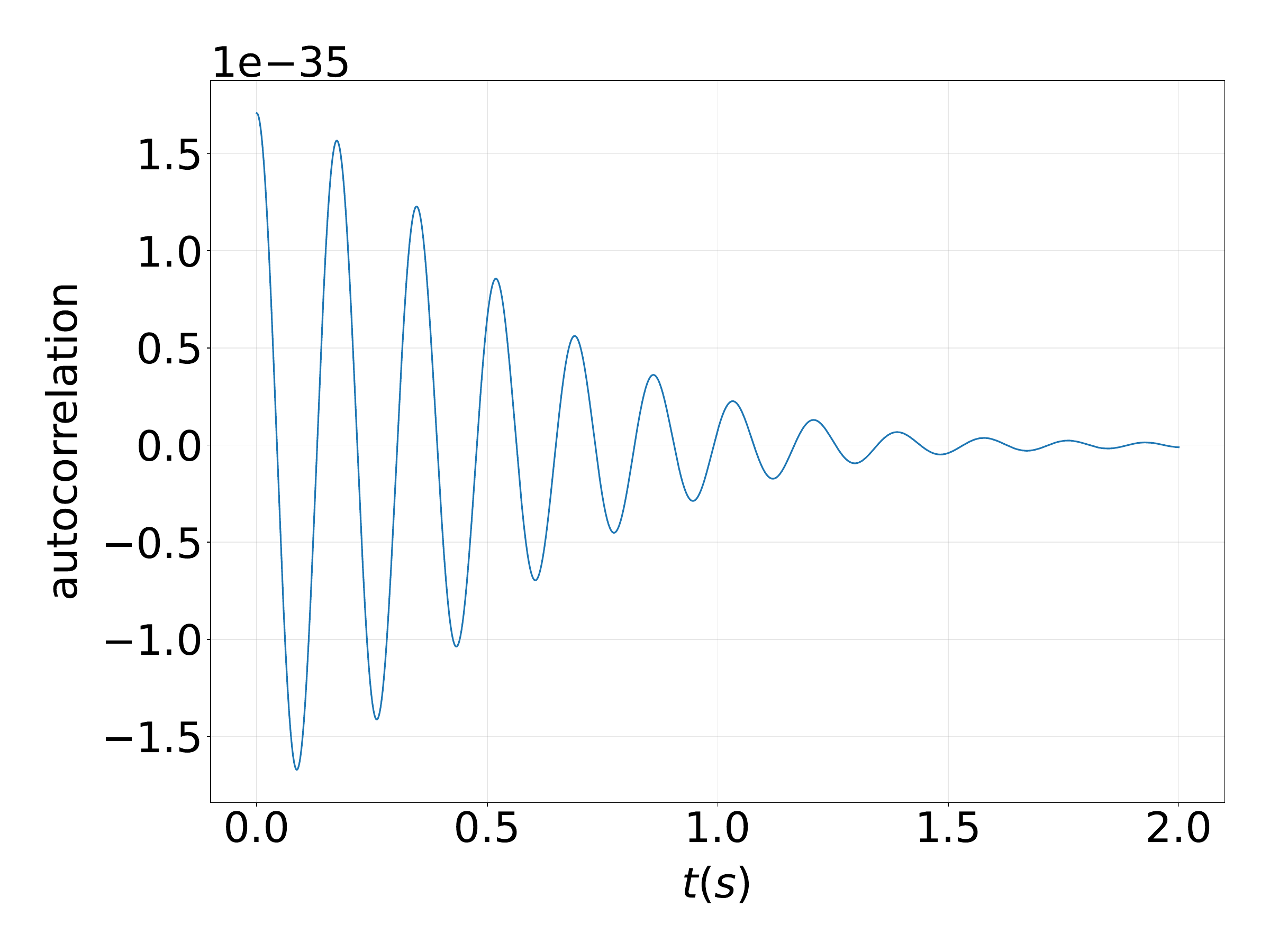}
    \includegraphics[width=0.49\linewidth]{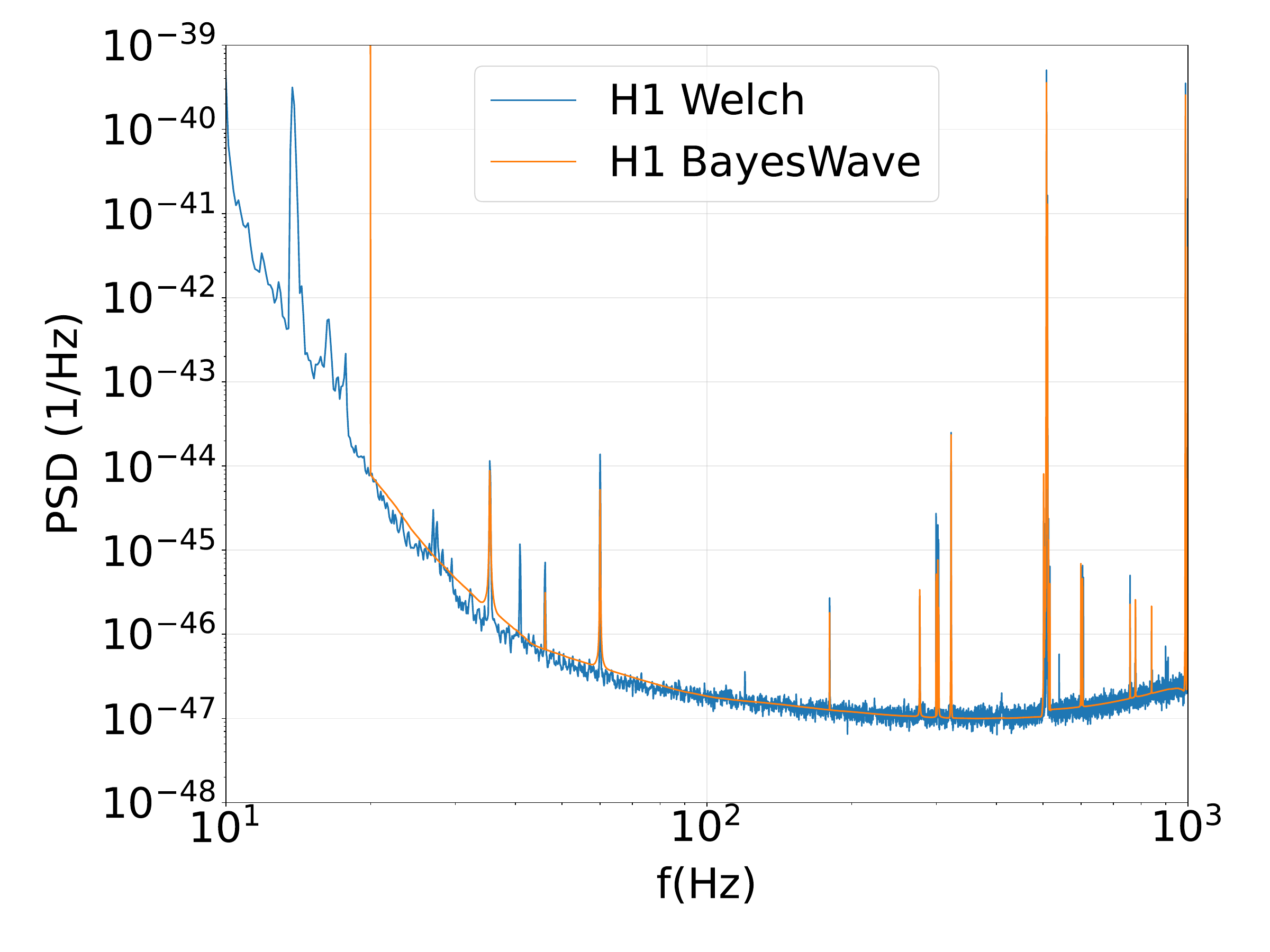}
    \caption{The estimated autocorrelation and PSD around the GW250114 event in the Hanford detector. The left panel shows the autocorrelation function, estimated by first computing the PSD using a media-averaged Welch method, which is shown in blue in the right panel. The band-limited BayesWave estimated PSD used in the FD analysis is shown in orange.)}
    \label{fig:acorr}
\end{figure*}
We outline the major implementation details of the code here.
\begin{enumerate}[leftmargin=*,itemsep=0.2em]
    \item \textbf{Time Segments}: Say one wishes to analyze the data segment between the sample indices $i_s \leq j \leq i_e$. In the time domain, this can easily be achieved by first constructing a truncated autocorrelation function with $i_e-i_s$ elements:
    \begin{equation}
        \rho_s = \rho(j\leq i_e - i_s)
    \end{equation}
    And then constructing from $\rho_s$ a noise covariance matrix $\cov_s$. The same likelihood form as in \eqref{eq:td_like_whitened} with the new covariance matrix will be the appropriate likelihood for the subset of the signal.
  \item \textbf{Data gaps:} If one needs to exclude a segment of data between $i_s \leq j \leq i_e$, one can set the data at these indices to zero. This can be equivalently achieved by deleting the corresponding entries in the autocorrelation function:
  \begin{equation}
      \rho_s = \{ \rho_j \,\, | \,\, j\notin [i_s \leq i_e] \}
  \end{equation}
  And constructing a covariance matrix from it. This ensures better memory traffic and computing. The rest of the analysis proceeds as before. 
  \item \textbf{Multi-detector likelihood}. The time-domain likelihood model for each detector is initialized as an instance of the `GWTransientTD` detector class, which encapsulates the data, the whitening operator, methods for waveform generation, and an interface to stochastic samplers such as bilby and dynesty. Multiple detectors are handled by a child class of the `GWTransientTD` class, which is also composed of $n_d$ instances of the `GWTransientTD' class, one for each detector. Each call to the multi-detector instance's likelihood method iteratively calls the corresponding functions of the individual detectors, and the log-likelihoods are summed. There is also a consistency test infrastructure to assess signal consistency across different time segments of the waveforms. We point the reader to ~\cite{prasad_msct} for details. The basic class structure of \emph{tdanalysis} is summarized in Fig.~\ref{fig:classd}.

  \item \textbf{Waveform generation}. Interface exists in the `GWTrantientTD` class to call waveforms from either `PyCBC` ~\cite{pycbc, Usman2016PyCBC} or `lalsimulation`~\cite{LVK2018LALSuite} in the time domain, using wrappers implemented in `waveformtools` ~\cite{waveformtools}.
  \item \textbf{Waveform conditioning}. Given the input parameters and analysis duration, waveforms are conditioned and injected into the detector frame by adding zeroes to the ends of the template to position the polarization peak at the given geocentric time. 

  In \emph{tdanalysis}, the time-of-arrival of the signal is parameterized differently from current implementations in PE codes like bilby~\cite{Ashton:2018jfp} and pycbc~\cite{Biwer2019PyCBCInference}. Currently, waveform models have their own definitions of coalescence time, which are usually based on a quantity (like frequency or amplitude of certain modes) in a gauge-fixed, co-precessing frame of the binary. However, these definitions vary from model to model. In \emph{tdanalysis}, I adopt an observer-based, mode-independent definition of the coalescence time, as the GPS time of the peak of the polarization of the signal before detector projection. Note that, depending on the polarization content, the direction of arrival of the signal relative to the detector, and the antenna patterns, the temporal location of the peak of the signal recorded in the detector may differ from the peak of the polarization. However, either is self-consistent, with the polarization peak being an observable, more gauge and model-independent.
  
  \item \textbf{Construction of whitening operator}. `tdanalysis` provides functionality to compute the autocorrelation function either directly by correlating a stretch of data devoid of signals, or by computing a Welch PSD and computing the inverse FFT. Fig.~\ref{fig:acorr} shows the autocorrelation function computed using the Welch method, using data around the GW250114 event. Given a whitening operator, the class `GWTransientTD` automatically has the ability to adapt $\bm{W}$ to the specified segment duration by recomputing the autocorrelation function and appropriately truncating it \cite{Levinson1947,Durbin1960}.
  \item \textbf{Computation of SNRs}. To mimic the Matched Filtered Signal-to-Noise ratios (MFSNR) in the frequency domain, we define the MFSNR in the time domain  as:
  \begin{equation}
      \rho_{mf} = \dfrac{\langle d, h \rangle}{\sqrt{\langle h, h\rangle}}
  \end{equation}
  using the time domain inner products. We also compute another SNR, which we call the effective SNR, defined by $\rho_{eff}^2 = \langle d, d \rangle - \langle r, r \rangle$, where $r = d - h$ is the estimated residue in the data, given a template $h$.

  \item \textbf{Data fetch and General pipeline details}.
  We interface \emph{tdanalysis} with \emph{gwdatafind} and follow a configuration-style pipeline, where the user inputs analysis settings in a multi-section ini file, much like the FD GW data analysis software \emph{bilby}. A "job factory" is provided to conveniently start a new project from a template configuration file and modify only the required keys. The job factory automatically creates the executable and prepares a SLURM script ready for submission. The analysis routine is responsible for downloading data from gwdatafind servers, preparing the necessary whitening operators, and resolving appropriate codepaths for various operations. The priors are input by the user in the same format as in bilby. 
  \item \textbf{Diagnostic checks}
  Several diagnostic checks are carried out before and after sampling. E.g., we test the properties of the residues for Gaussianity, check for overwhitening of the data, quantify residual correlations, etc. Fig.~\ref{fig:wf} shows the whitened residue distribution in the Hanford detector after analyzing GW250114, the details of which can be found in Sec.~\ref{sec:results}.
\end{enumerate}
This implementation and choice of segmentation enables the analysis of any portion of the time domain signal, be it inspiral, ringdown, merger, or the whole signal, possibly excluding a portion in between, using the corresponding portion of a full IMR waveform.  Once the analysis segments are defined by their GPS times, the peak time of the signal determines the portion of the template waveform that lies in the analysis window. We note that this implementation is an improvement over existing implementations for analyzing individual segments of the signal in many ways. First, in existing time domain studies like ~\cite{Miller:2023ncs, Miller_2025, Miller:tdinf:2025}, some of the extrinsic parameters like sky locations, polarization, and peak time are fixed due to limitations of speed. The accelerations presented in this work enable sampling over all parameters, including the peak time, with judicious resources and in a feasible time. In fact, as described in ~\cite{prasad_msct}, we show that one can simultaneously analyze multiple segments of data, corresponding to a 25-dimensional parameter space, owing to the accelerations implemented here. As a corollary, the analysis of the ringdown segment can thus include non-linearities as well, with no ambiguities in the choice of start time, as it is also sampled over.

Related models that analyze the data using portions of the IMR waveforms have been used in the past for ringdown studies. In~\cite{Brito:2018rfr}, the post-merger part of the signal from a non-spinning, quasi-circular source is modeled through the corresponding portion of an effective one-body waveform, inferring the ringdown frequencies even though the full IMR signal is analyzed, fixing sky locations in some studies. In ~\cite{shear-news2020, MyThesis, Prasad:2023bwa}, the infalling radiation at the outer common dynamical horizon was fit to the post-peak portion of aligned-spin full IMR waveform to infer the parameters of the source, keeping the start time fixed at the peak of the $\ell=2, m=2$ mode. Later ~\cite{CalderonBustillo:2020rmh, Qiu:2023, Gennari:2023gmx} also use the post-peak portion of a full IMR waveform to analyze the post-merger signal in a similar fashion. The sky location and polarization angles are fixed in ~\cite{CalderonBustillo:2020rmh}. In ~\cite{Qiu:2023}, the start time and extrinsic parameters are sampled over, while in ~\cite{Gennari:2023gmx}, the sky location and peak time are fixed. More recently, ~\cite{Chandra:2025} revisits ~\cite{CalderonBustillo:2020rmh} using modern waveform models with multiple choices of start time, while keeping it fixed for each run. They fix the sky location and the merger time in each of their analysis. Although not addressing consistency across segments, ~\cite{Miller:2023ncs, Miller_2025} analyze individual segments of the waveform, including the ringdown portion, to extract precession physics. They fix the sky position, polarization, and the coalescence time in their analysis. However, this implementation allows all parameters to be feasibly sampled over, including the peak time of the signal feasibly. 

Secondly, like ~\cite{Miller:tdinf:2025}, this is a fully time domain pipeline with a Gaussian Likelihood, thus not needing the Whittle approximation.

\section{Cost analysis and Likelihood acceleration techniques for the time domain likelihood}
\label{sec:cost}

\subsection{Algorithmic acceleration and Complexity analysis}
\label{subsec:algo}
The process of parameter estimation using stochastic sampling via a Monte-Carlo technique involves repeated evaluations of the likelihood function in \eqref{eq:td_like_whitened}, sometimes exceeding $10^8$ evaluations before the samples converge sufficiently. In this subsection, we discuss ways to speed up this operation.

We begin this section by analyzing the major computational costs associated with each stage of the end-to-end time-domain PE pipeline, excluding waveform generation, from conditioning to detector response and likelihood evaluation. Then, we examine and discuss the optimal algorithms and computational strategies used in the current pipeline implementation and note those currently under development in Sec.~\ref{subsec:algo}. Then, we report on acceleration attempts by vector CPUs and GPUs and discuss their limitations in Sec.~\ref{subsec:hardw}. Finally, we briefly mention the efforts in software acceleration that led to an optimized pipeline with end-to-end optimizations.

As discussed earlier in Sec.~\ref{sec:likelihood}, it is necessary to consider both the memory traffic cost and the computational complexity, as the former may pose a bottleneck, resulting in less than optimal use of the hardware. We carry out a cost analysis in this light.

\subsubsection{Data projection and conditioning}

Once the polarizations ($h_{+, x}$) and the antenna patterns $f_{p, x}$ of the detectors are generated, the polarizations are projected onto the detector frame by computing the dot product $h_+ f_+ + h_x f_x$. This is an $O(N)$ operation, both in terms of computational complexity and memory traffic.

Conditioning the waveform in the time domain involves placing the projected waveform at the appropriate position in the data vector, based on the signal's geocentric time. This operation has $O(N)$ computational complexity and $ T(M)$ memory traffic.

\subsubsection{Factorization of the Covariance matrix} 

As discussed in Sec.~\ref{sec:likelihood}, one can factorize the covariance matrix using Cholesky decomposition $\cov = \cholL^{\top} \cholL$, endowing the analysis with a whitening operator. This operation needs to be done once, and for a dense $\cov$, the complexity is $O(N^3)$; for Toeplitz/banded matrices, some implementations reduce this to $O(N^2)$. Either way, the result can be cached, and this is a one-time computational cost, making a small difference to the total runtime, provided the signals are O(1) seconds long.  

Once $\cov^{-1}$ is precomputed, without the use of factorization, the evaluation of the likelihood involves a computational complexity of $O(N^2)$. The corresponding memory traffic constitutes streaming all the elements of the matrix and the vectors, and thus totals to $T(aN^2 + bN)$ at leading order, where $a,b$ are some constants

If instead one carries out a Cholesky decomposition, the likelihood via Sec.~\ref{sec:likelihood} operation i.e. explicit GEMV~\cite{LawsonEtAl1979BLAS, DongarraEtAl1988Level2BLAS} of $\cholL\cholL^{-1} \hvec$ is still $O(N^2)$ and $T(N^2)$, but the constants pre-multiplying them are approximately halved, because $L$ is lower triangular and half of the elements are zero. Thus, the total theoretical speedup is 2x. We call this method the Cholesky Decomposed Operator (CDO) method for brevity.

Given this decomposition, there are then two approaches to the evaluation of the likelihood in ~\eqref{eqn:lik}. One involves TEMV~\cite{Gray2006ToeplitzCirculantReview, ChanJin2007IterativeToeplitzSolvers}, solving for $\bar{\bm{r}} = lower\_triangular(\cholL, \bm{r})$ $O(N^2 + N)$. We use `scipy.linalg.solve\_triangular` wrapper for this purpose. We call this the `CDL` method. The other method is the one described previously. This involves pre-computing and caching $\cholL^{-1}$, and using GEMV to obtain $\bar{v}$ i.e. $\bar{\bm{v}} = cholL^{-1} \bm{v}$ ($O(N^2)$). Although the latter is better in the sense of throughput due to the full use of Fused Multiply Add (FMA) and Advanced Vector Extensions (AVX)~\cite{IEEE754-2008,IntelOptRefManual,IntelSDMVol2A,Lomont2011AVXIntro,NvidiaFloatingPointIEEE754,AgnerFogInstructionTables}, the former is more numerically stable (avoids round-off errors and potential underflows and overflows during inversion). However, we find that the $CDL$ method is not feasible in practice because it may require many iterations to converge, which adds a linear factor to the complexity and outweighs the 2x speedup from the lower triangular structure.

In the current `CDO' implementation, we pre-compute and cache $L$ and the whitened data vector $\bar{d} = \bm{W} \ddat$ once at initialization. For repeated evaluations, we use the `GEMV` approach directly. The likelihood is then finally computed by a simple vector dot product of the whitened residue, $\bar{\ddat} - \bar{\hvec}$. 

The overall computational cost in the likelihood evaluation in the current `CDO` implementation is therefore $O(aN^2 + bN)$. 

Both methods are nevertheless available in \emph{tdanalysis}.

\subsubsection{\textbf{Gohlberg - Semencul theorem and Convolution / \FFT{}-based method:}}  
Given that the `GEMV` step is memory-bandwidth-limited and underutilizes the computational capabilities, we sought to reduce memory traffic. This, we reasoned, was possible only if the inner product could be evaluated without streaming the matrices into the CPU caches from memory. This meant a hunt to reduce the `GEMV` operation into a collection of vector algebra operations.

For evaluating inner products of the form $\mathbf{a}^T Q \mathbf{b}$ when $\mathbf{C}$ is circulant, instead of using matrix operations to compute $\bar{\bm{a}} = \mathbf{W}^{-1} \mathbf{a}$ with $W$ being am approriate factor of $Q$, which is by far the most expensive step in likelihood evaluations, one can embed the $Q$ Toeplitz matrices into circulant matrices at the cost of doubling the vector length by stitching the ends of the data vector together, bringing artificial periodicity. Since circulant matrices are diagonalized in the Fourier basis, one can then use FFT algorithms to essentially compute \eqref{eq:td_like_whitened} in place of \eqref{eq:td_like_whitened}, with a corresponding `PSD' derived from the doubled Toeplitz matrix. The computational cost of this procedure scales as $O(2N log(2N)) + 2N$, which is particularly advantageous when $N$ is very large. In contrast to the time-domain methods described above, the memory traffic for FFT-based approaches is only $T(N)$, since only multiple vectors need to be streamed into memory rather than matrices, making the problem more compute-bound. This is the biggest advantage of this method.

However, the evaluation of the inner product in \eqref{eq:td_innerprod_whitened} involves the inverse of a Toeplitz matrix. While the Covariance matrix is a symmetric Toeplitz matrix, the inverse matrix is generally not. This prevents a direct application of the circulant embedding trick to achieve O(NlogN) scaling. However, the Gohberg-Semencul (GS) theorem comes to the rescue \cite{GohbergSemencul1972,Trench1964,Gray2006Toeplitz}. The GS theorem provides a way to decompose the inverse of a Toeplitz matrix as a sum or difference of the products of two Toeplitz matrices.
\begin{equation}
\begin{aligned}
\Cbf^{-1}
&=
\frac{1}{x_1}
\left(
\Lbf_x \Lbf_x^{T}
-
\Rbf_x^{T} \Rbf_x
\right), \\
\Cbf\,\mathbf{x}
&=
\mathbf{e}_1,
\qquad
\mathbf{x}=(x_1,\dots,x_n)^T .
\end{aligned}
\end{equation}
$\Lbf_x$ is the lower-triangular Toeplitz matrix with the first column
$\mathbf{x}=(x_1,\dots,x_n)^T$, and $\Rbf_x$ is the strictly upper-triangular
Toeplitz matrix with first row $(0,x_n,x_{n-1},\dots,x_2)$.
This means that evaluating the inner products in \eqref{eq:td_innerprod_whitened}, which corresponds to evaluating quadratic forms, involves the action of Toeplitz matrices on vectors, which can be circulantly embedded by doubling the lengths of the data vectors. For a Toeplitz matrix $\Tbf\in\mathbb{R}^{n\times n}$ and vector $\mathbf{x}\in\mathbb{R}^n$,
let $\mathbf{S} \in\mathbb{R}^{m\times m}$ be a circulant embedding of $\Tbf$ with $m\ge 2n-1$. 
If $c$ is the first column of $\mathbf{S}$, and $\tilde{\mathbf{x}}\in\mathbb{R}^m$
is $\textbf{x}$ zero-padded to length $m$, then
\begin{align}
\Tbf x
&= \mathcal{P}_n\,\mathbf{S}\,\tilde{x}, \\
&= \mathcal{P}_n\,\mathcal{F}_m^{-1}
\!\left[
\mathcal{F}_m(c)\odot
\mathcal{F}_m(\tilde{x})
\right],
\label{eq:toeplitz_fft}
\end{align}
where $\mathcal{P}_n$ restricts to the first $n$ components, \(\mathcal{F}_m\) is the
length-\(m\) discrete Fourier transform, and \(\odot\) denotes elementwise
multiplication.

This accelerates the evaluation of the quadratic forms to O(N log (N)) as FFT algorithms can be used to operate circulant Toeplitz matrices on data vectors. We call this the Gohberg-Semencul Circulant Embedding (GSCE) method. We exploit this theorem and find that this is the fastest algorithm one can use to speed up likelihood evaluations, while maintaining exactness, i.e., without the need for circulant approximations.

\subsubsection{\textbf{Reduced-basis compressions:}} 

Even when $\cov$ is not circulant, given its symmetric structure, it admits a complete set of $N$ orthonormal eigen-basis vectors. In practice, all the principal components of $\cov$ do not contribute equally to the evaluation of the Likelihood. Therefore, it is possible to construct a reduced-order basis representation of the whitened noise, data, and signal by Singular Value Decomposition (SVD) and retain only the necessary $M$ number of basis vectors. This will reduce the cost of repeated inner-product evaluations, complementing efforts in frequency-domain speedups such as reduced-order quadrature \cite{Qi2021ROQ} and reduced-basis filtering \cite{Dhurkunde2022ReducedBasisMF}.  Such an approach, although it will not alter the dominant power dependence on complexity, reduces it to $O(M^2)$, with $M\leq N$, thereby also decreasing memory traffic. This will be presented elsewhere.

To summarize this section, various speedup methods improve overall scaling in computational complexity/ memory traffic, yielding a factor of 2 improvement at best, but do not change the overall complexity, which remains at $O(N^2)$ and $ O (MN^2)$. Only the convolutional FFT-based implementations offer superior scaling similar to native FD methods, at the cost of doubling the effective sample size. In this next subsection, we explore how the effective use of hardware could help speed up these algorithms.

\subsection{Hardware acceleration}
\label{subsec:hardw}
The adaptability of time domain algorithms that are $O(N^2)$ computationally expensive, as opposed to FFT-based methods that are $O(N \log  N)$, is primarily due to advancements in computational hardware. This involves enhancements to modern CPU, GPU, and memory over the past two decades. We describe here the specifics of how these improvements enable time-domain methods to be adapted for gravitational-wave astronomy.

\subsubsection{Memory bandwidth}
Consider analyzing a signal that is $4s$ long, sampled at $ 4096 Hz$. This corresponds to $N=16384$. Given the sensitivities of the current generation of detectors, FP32 is sufficient to avoid numerical underflow/overflow in various matrix operations during likelihood evaluation, given the inverse covariance or the whitening matrix. This means that the size of any $N\times N$ matrix in memory (e.g., the covariance matrix)is $1GiB$. Modern RAMs have a memory bandwidth of around $50-90$ GB/s at dual data rate. This implies that, as this large a matrix cannot fit in any current CPU cache, during each iteration of likelihood evaluation, a $1$GiB matrix (and vectors of $16$KiB) needs to be streamed, which would take ~$T(10^{1})$ms. 

Compare it with the hardware available in e.g., 2005, when memory bandwidth was 100 times slower, meaning each likelihood evaluation would take at least $T(1)$s. Even when the first GW event, GW150914, was detected, the memory bandwidth was around $10$ GBps, so each likelihood evaluation would still take a second, making it infeasible for stochastic sampling.

\subsubsection{CPU acceleration}

Various improvements in CPU instruction architectures, including Advanced Vector Extensions (AVX) ~\cite{IntelOptRefManual,IntelSDMVol2A,Lomont2011AVXIntro}, Fused Multiply-Add (FMA) ~\cite{IEEE754-2008, NvidiaFloatingPointIEEE754, AgnerFogInstructionTables}, and increased core counts, and their use via directives such as OpenMP~\cite{DagumMenon1998OpenMP, OpenMPARB2021OpenMP52} and POSIX/Pthread~\cite{IEEE1003.1c-1995Pthreads, Butenhof1997Pthreads}, have greatly enhanced CPU vector capabilities. 

With AVX, Single Instruction Multiple Data (SIMD) means that processors are no longer restricted to execute a single floating-point instruction at a time. They are now equipped with registers that allow a single instruction to be applied to multiple floating-point (and integer) data. The registers on modern processors can hold upto 8 FP64 values or 16 FP32 values.

With FMAs, multiple FP instructions, such as multiply and add, can be combined and executed together, leading to faster execution and lower roundoff error.

Shared-memory models such as OpenMP/threads allow harnessing the many cores in CPUs of this era to operate on large datasets, thereby benefiting linear algebra operations. Effective use of these capabilities is enabled by BLAS~\cite{LawsonEtAl1979BLAS, DongarraEtAl1988Level2BLAS} and LAPACK~\cite{AndersonEtAl1999LAPACKUsersGuide} implementations of linear algebra routines that are being continually improved to adapt to modern hardware, which Python's `numpy` and `scipy` use in their backend. 

These improvements have enabled modern CPUs to achieve significantly high FLOP/s. At $1$ GFLOP/s, a GEMV would take about $500$ ms. Modern CPUs are capable of handling up upto $10^3$ GFLOP/s, which translates to $0.5$ ms, much smaller than the memory traffic time. Thus, the likelihood evaluation operation is predominantly memory-bandwidth-limited, and we cannot fully exploit modern CPUs' capabilities.

The nested sampling procedure is inherently serial, requiring likelihood comparisons to decide whether to accept the proposed samples. This severely restricts the use of a CPU's parallel computing capabilities. The only way to parallelize the sampling process is through the proposal step, where new samples are drawn from the prior and presented for acceptance. Multiple threads/processes can work independently to draw samples from the prior and pre-compute the likelihoods, which are then stored in a pool. The stepper then has access to many candidates for accepting new samples as live points, speeding up the sampling process. This can be achieved through process-based parallelism, e.g., `multiprocessing`~\cite{PEP371Multiprocessing, PythonDocsMultiprocessing}. Many nested and Markov-Chain-based samplers like 
`dynesty`~\cite{Speagle:2019ivv}, `Bilby`~\cite{Ashton:2018jfp}, `emcee`~\cite{ForemanMackeyEtAl2013emcee}, `PTMCMC`~\cite{VousdenFarrMandel2016PT},
`Zeus '~\cite {KaramanisBeutlerPeacock2021zeus}, `UltraNest '~\cite {Buchner2021UltraNest}, etc., exist today that provide the functionality to use a multiprocessing pool to achieve this.

At first, a process-based parallelization seems to compete with OpenMP for the same CPU resources, potentially slowing down linear algebra operations during individual likelihood evaluations and waveform generation. However, as the most expensive step in the likelihood evaluation is whitening/evaluation of the quadratic form, which is memory-bandwidth-limited, we show in Sec.~\ref{sec:bench} that OpenMP is not effective at using all the compute power available across cores for dense matrix computations. Therefore, we find that using process-based parallel constructs, such as Python's `multiprocessing`, accelerates the sampling process. In Sec.~\ref{sec:bench}, we will discuss the interplay between OpenMP and multiprocessing.

\subsubsection{GPU acceleration}

General Purpose Graphics Processing Units are a powerhouse for linear algebra, exploiting the Single Instruction Multiple Threads (SIMT) nature of linear algebra routines. Their architecture, composed of hundreds and thousands of streaming multiprocessors, is perfect for operating on large matrices. Modern GPUs can have a throughput of up to 1 TFLOP/s.

GPUs not only offer high throughput but also have always had higher memory bandwidth than CPU RAM, due to their original requirement for real-time graphical rendering. Modern-day GPUs have a memory bandwidth of up to 1 TB/s, more than 10 times that of RAM. This makes them perfectly suited for time-domain analysis of long-duration signals. 


GPU memory, on the other hand, has also been slowly rising in capacity, reaching up upto 100GB on data center GPUs and a few GBs on personal machines. This allows matrices of the order $1-2GBs$ to be stored on VRAMs, which was infeasible to handle a decade ago on any computing platform. The bandwidth dictates that streaming such matrices takes less than $2ms$, suggesting the same bottleneck, even with GPUs' improved memory bandwidth. However, on GPUs, there exists another overhead: GPU kernel startup cost. Kernel latency is usually constant, independent of the operation being performed, and is of the order of a few tens of $ms$. Thus, given the requirement for repeated evaluations of the likelihood, and the fact that the template vectors need to be transferred from the CPU at each evaluation, the Kernel latency dictates the speed of evaluations on GPUs, given that the whitening operator $\bm{W}$ and the whitened data vector are cached on VRAM.
\begin{figure}
    \centering
    \includegraphics[width=\linewidth]{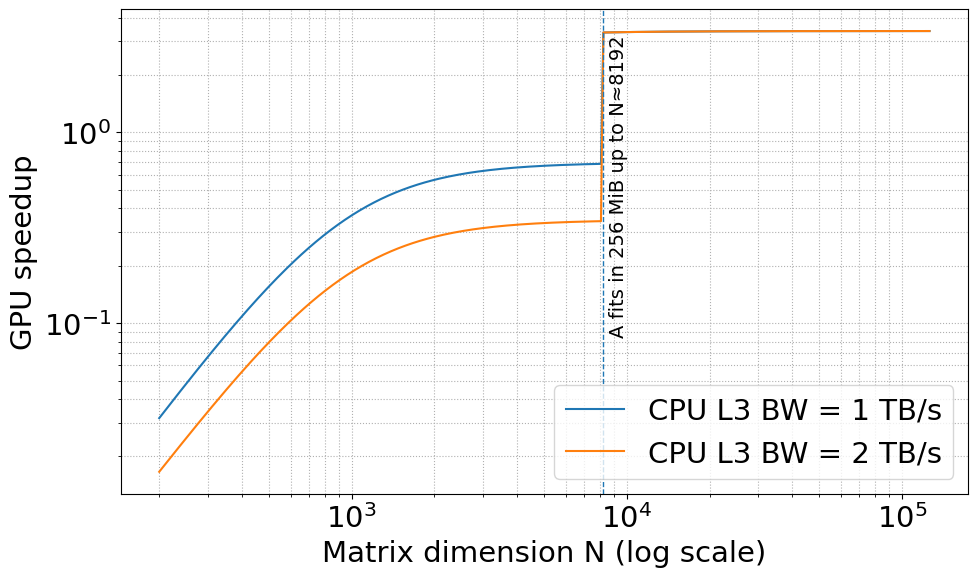}
    \caption{Caption}
    \label{fig:gpu_spdup}
\end{figure}
Despite this bottleneck, GPUs offer noticeable speedups over CPUs, especially when $N$ is large. To see this, consider the dense GEMV operation on a typical 64-core CPU. Since the operation is predominantly memory bandwidth limited, the time taken can be estimated as
\begin{equation}
    t = \dfrac{8N^2}{B}
\end{equation}
Where the bandwidth is $ B\approx1$ TBps on GPUs like the Nvidia A100 and AMD MI100. This means that, in order to be slower than the GPU kernel startup time of $10ms$, one would need to have $N\approx40,000$, which corresponds to about $5.5$s at a sampling rate of $4$kHz. From this point onwards, the total GPU time will be dominated by memory traffic. Thus, for small $N\approx 1000$, the total GPU time is unaffected due to the approximate constant kernel startup time, producing an approximate GPU speedup like the one shown in Fig.~\ref{fig:gpu_spdup}. Note that if the matrices fit completely in the CPU's L3 cache  (which are large in modern times, and assumed to be about $256MiB$ as in many current-day AMD EPYC processors in this plot), and can be kept hot, the CPU could be faster at likelihood evaluations because the L3 caches could be larger. As $N$ increases and the matrices cannot fit on caches, the speedup is approximately constant for GPUs, since the bandwidth ratio determines the total time per evaluation. Since GPU VRAMs are 3-5x faster, that speedup persists throughout, as the operation is bandwidth-limited.

The takeaway from this analysis is that GPUs can achieve up upto 5-10X speedups in likelihood evaluation, primarily due to memory bandwidth improvements. In the current implementation of \emph{tdanalysis}, we provide an option to offload the heaviest linear algebra computations on the GPU in the `CDO` method. The whitening operator and the data vector are cached on GPU VRAM, and the whitened data vector is precomputed once and cached. At every likelihood call, the CPU transfers a template vector $\hvec$ to the GPU, which whitens it using GEMV, and the result is transferred back to the CPU. The likelihood evaluation is completed on the CPU by computing the residue and a dot product.

We also tested performing the remaining operations, i.e., computing the residue and dot products, on the GPU itself within the same kernel, thus returning a single float to the host. However, since it scales as $O(N)$, we found that it does not provide noticeable speedups, at least for the moderate values of $N$ we tested.

As we shall see later, the `GSCE` method implemented on CPUs offers the highest speedups among the algorithms considered here. The GPU implementation is currently in testing, and we find it beneficial for evaluating long-duration $N\gtrapprox 10^7$ signals. For products that run on 4s of $4KHz$ data ($ N\approx 10^4$), we use CPUs.

\subsection{Software acceleration}

\subsubsection{Memory ordering}
In Python, vectors and matrices are not guaranteed to be contiguous in memory. We find that ensuring all vectors and matrices are contiguous in memory, and in Fortran ordering, can yield up upto 2x speedups, which we implement.

\subsubsection{Code optimization}

Most of the other software speedups we implement are relevant to GPUs. When offloading linear algebra or inner-product computations to GPUs using the `CDO` method, there are multiple Application Programming Interface (API) options. The most common and generic software API for this is TensorFlow~\cite{AbadiEtAl2016TensorFlow} or PyTorch~\cite{PaszkeEtAl2019PyTorch}. The other option is CuPy~\cite{OkutaEtAl2017CuPy}, which only currently works on Nvidia GPUs. We find that CuPy offers upto 2x speedup when compared to TensorFlow codepaths. Drawing on this inspiration, we implement a custom HIP~\cite{AMD2021HIPProgrammingGuide} kernel to offload the whitening operation using the `CDO` method to AMD GPUs, and compile it with hardware-specific optimizations. We achieve up upto 1.5x speedups compared to TensorFlow. \emph{tdanalysis} provides users with the choice to select the software backend for offloading computations to the GPU.

\subsubsection{Precision}

We note that the autocorrelation function generally has a dynamic floating-point range of $10^{-42}$ to $10^{-34}$, i.e., spanning about 8 orders of magnitude (see Fig~\ref{fig:acorr}). Given that FP32 has a dynamic range of $10^{\pm 38}$, it is possible to carry out all the analysis with FP32 with appropriate normalization. However, the necessary matrices and whitening operators should be precomputed in FP64 to avoid numerical underflow or overflow. This generally speeds up calculations by a factor of 2, as CPU vector registers (such as AVX and SSE) can hold twice as much data, enabling higher throughput. $\emph{tdanalysis}$ provides users with the choice of precision for likelihood evaluations.

\section{Benchmarks}
\label{sec:bench}
In this section, we discuss preliminary results and benchmarks for the current version of `tdanalysis`'s CPU implementation. 

The hardware used for production runs is as follows:
\begin{enumerate}
    \item \textbf{CPU/ Host}. CPU-based runs and the host CPU used for GPU runs are performed on an AMD EPYC 7763 dual-socket system with 64x2 cores. They have an L3 cache of 256 MiB. The peak theoretical memory bandwidth of the dual socket 16-channel DDR4-3200 RAM is around $400$ GB/s.
    \item \textbf{GPU}. For GPU computing, we test on both NVIDIA A100 and AMD MI 100 GPUs. We use MI100 for extensive runs as it offers almost the same performance at a lower resource cost.
\end{enumerate}
The software environment is as follows:
\begin{enumerate}
    \item \textbf{Stochastic sampling}. We use the \Bilby code to sample the posteriors. We use Nested sampling with the `dynesty` back-end and \Bilby's acceptance walk front-end for new sample proposals.
    \item \textbf{Linear algebra}. On CPUs, all linear algebra is handled using Python's `scipy'~\cite{Jones_SciPy} and `numpy'~\cite{van_der_Walt_NumPy} modules, which use implementations from BLAS and LAPACK for linear algebra operations, and are parallelized with OpenMP. 

    On GPUs, we provide the options to use \emph{torch/TensorFlow} back-end, which works generically with both AMD and Nvidia GPUs. Alternatively, we provide the `CuPy' interface that performs faster than \emph{torch/TensorFlow} back-ends on NVIDIA devices. But they don't work on AMD GPUs yet. As a third option, for AMD GPUs, we provide a custom HIP kernel for the GEMV process with SWIG bindings that can be compiled with native arch optimizations for the specific AMD GPU at hand. This works noticeably faster than the torch backend on AMD GPUs.
\end{enumerate}

The test runs are performed by injecting a zero-noise GW event with parameters similar to GW250114~\cite{LIGOScientific:2025rid, LIGOScientific:2025obp} and GW230814~\cite{LIGOScientific:2025cmm}. The latter is a single-detector event, allowing a full PE run with slower algorithms considered here, while keeping computational cost relatively low, as it involves a single likelihood call per evaluation. We run the PE on i). I) CPU with varying sizes of multiprocessing pools, ii) one GPU (MI100) with no multiprocessing pool. We report the results in the next Section

\section{Results}
\label{sec:results}
\subsection{Performance and scaling}
\begin{figure*}
    \centering    
\includegraphics[width=0.32\linewidth]{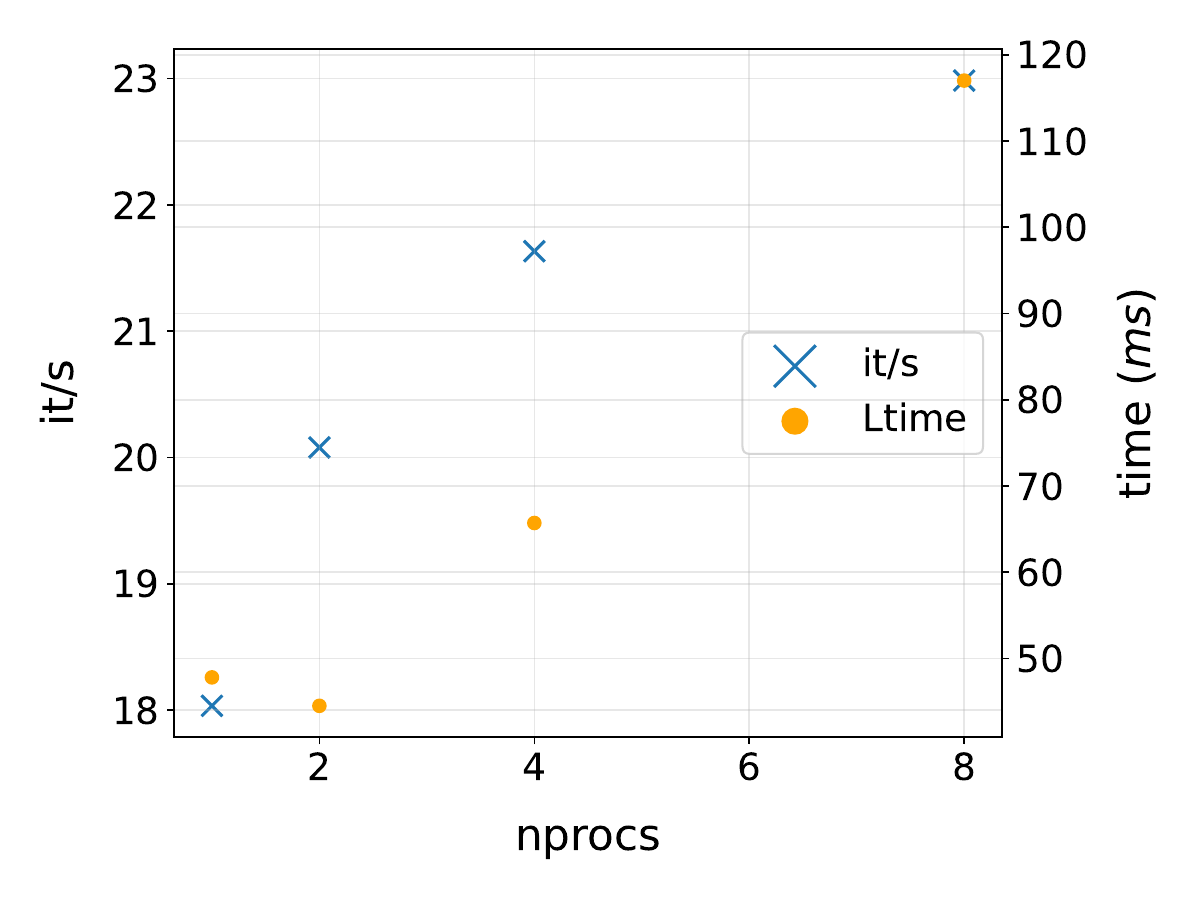}
    \includegraphics[width=0.32\linewidth]{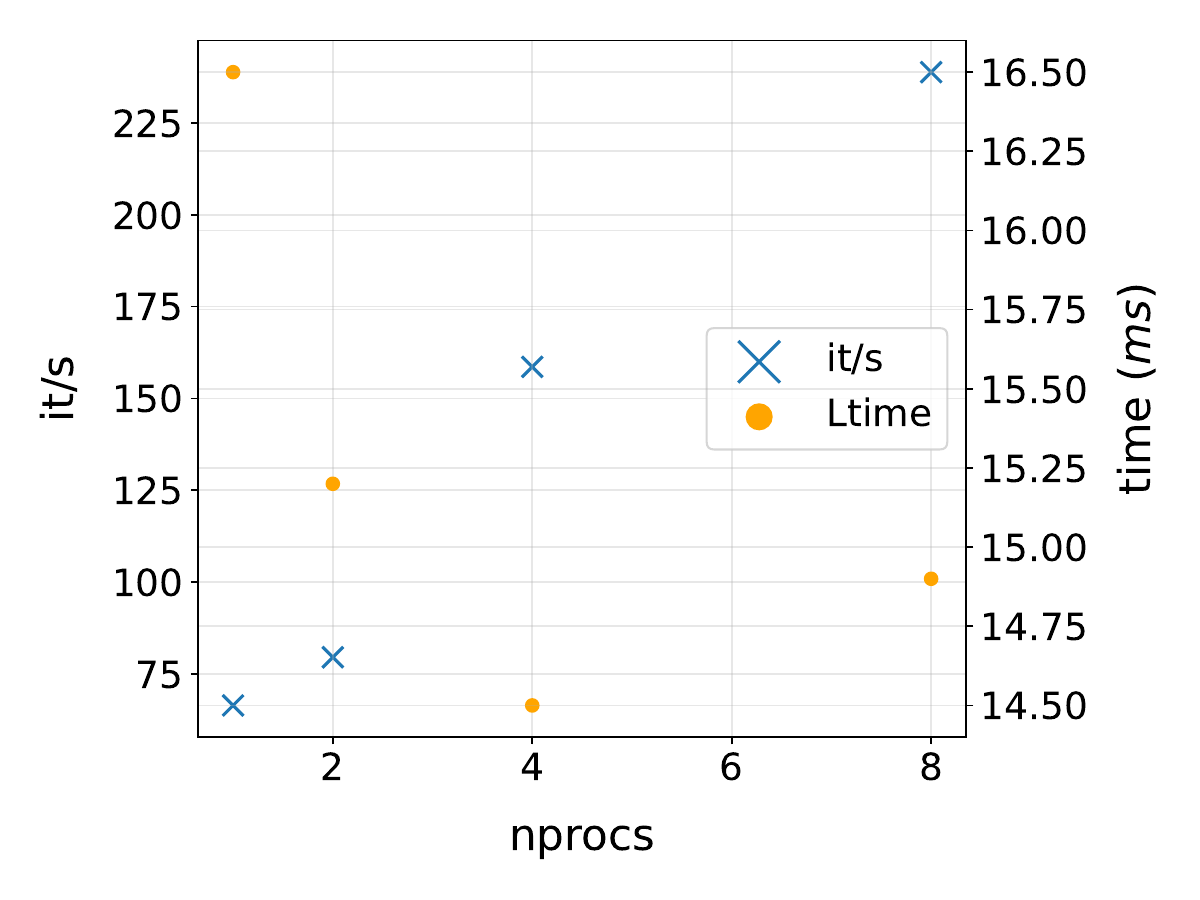}
    \includegraphics[width=0.32\linewidth]{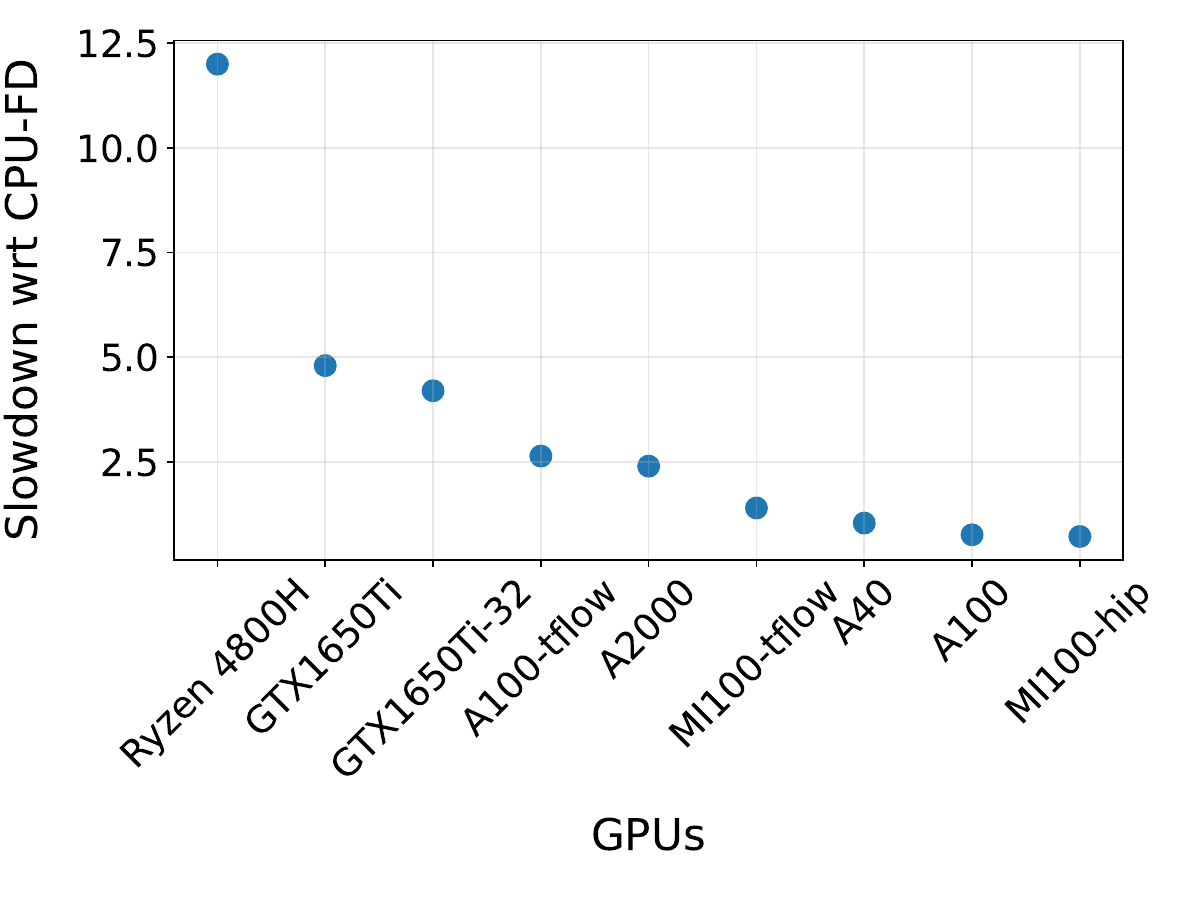}
    \caption{Benchmark results for direct Cholesky decomposed whitening operator method, and the GSCE method. The left plot shows the combined number of likelihood calls per second of the sampler across the multiprocessing pool on the left axis, and the single likelihood evaluation time on the right axis. The plot in the center shows the performance of the GSCE method. The right panel shows the ratio of likelihood evaluation time on the device specified on the x-axis to that of a frequency domain likelihood evaluation on a CPU.}
    \label{fig:bench1}
\end{figure*}
We compare the performance of two methods of evaluating the likelihoods i). Using the `CDO' method on CPUs and GPUs, and iii) Using `GSCE' on GPUs. All benchmarks are run on an Intel 258v processor with a memory bandwidth of 8533 MT/s. The device remained plugged into the power socket. The benchmarks are carried out by running the bilby sampler with different numbers of multiprocessing pools, from 1 to 8, to examine the interplay between process-based parallelism and shared memory parallelism. At each multiprocessing pool value, the OMP\_NUM\_THREADS value is set such that OMP\_NUM\_THREADS $\times npool = 8$, the number of physical cores on the device. The results are shown in Fig.~\ref{fig:bench1}. 

We do not benchmark the lower-triangular solve method because likelihood computation times are unpredictable in this approach. The GPU benchmarks are shown as single likelihood evaluation times in Fig.~\ref{fig:bench1}, as a multiprocessing pool is not currently usable with GPUs. However, we are currently working on a solution to remedy this.

The benchmark results can be summarized as follows.
\begin{enumerate}
    \item Single likelihood evaluation times on CPUs are about 3-15 times slower than frequency domain when directly using the `CDO' method. When using a multiprocessing pool, the average number of likelihood calls per second is approximately 10 times slower.
    \item On GPUs, the `CDO' method attains about 2-10x speedup per likelihood evaluation over CPUs. However, they are still 1-5x slower than the FD method. Using data center GPUs capable of FP64 compute and optimizing custom GPU kernels with HIP can match CPU FD valuation times. 
    \item When using `GSCE' on CPUs, the likelihood evaluation times and the average number of likelihood calls in a given duration across the multiprocessing pools are approximately the same as in the FD approach when using this method. 
    
    Implementing and benchmarking this method, we find that the GSCE method is approximately only slower than the FD method by a factor of $\sim2$. Thus, it is possible to do TD analysis with only twice the computing resources as the FD methods.
    
    \item On CPUs, the `GSCE' method is roughly 10 times faster than the `CDO' method, while at the same time not using as much memory bandwidth as the `CDO' method, leaving room for process-based parallelization. The effective speedup is thus $\approx10\times n_{pool}$, where $n_{pool}$ is the pool size for proposals.
    \item We find that using process-based parallelism is more advantageous than shared-memory parallelism. Thus, we recommend using a multiprocessing pool size equal to the number of physical cores available to achieve the best results.
\end{enumerate}
\subsection{Parameter estimation}
\begin{figure*}
    \centering
    \includegraphics[width=\linewidth]{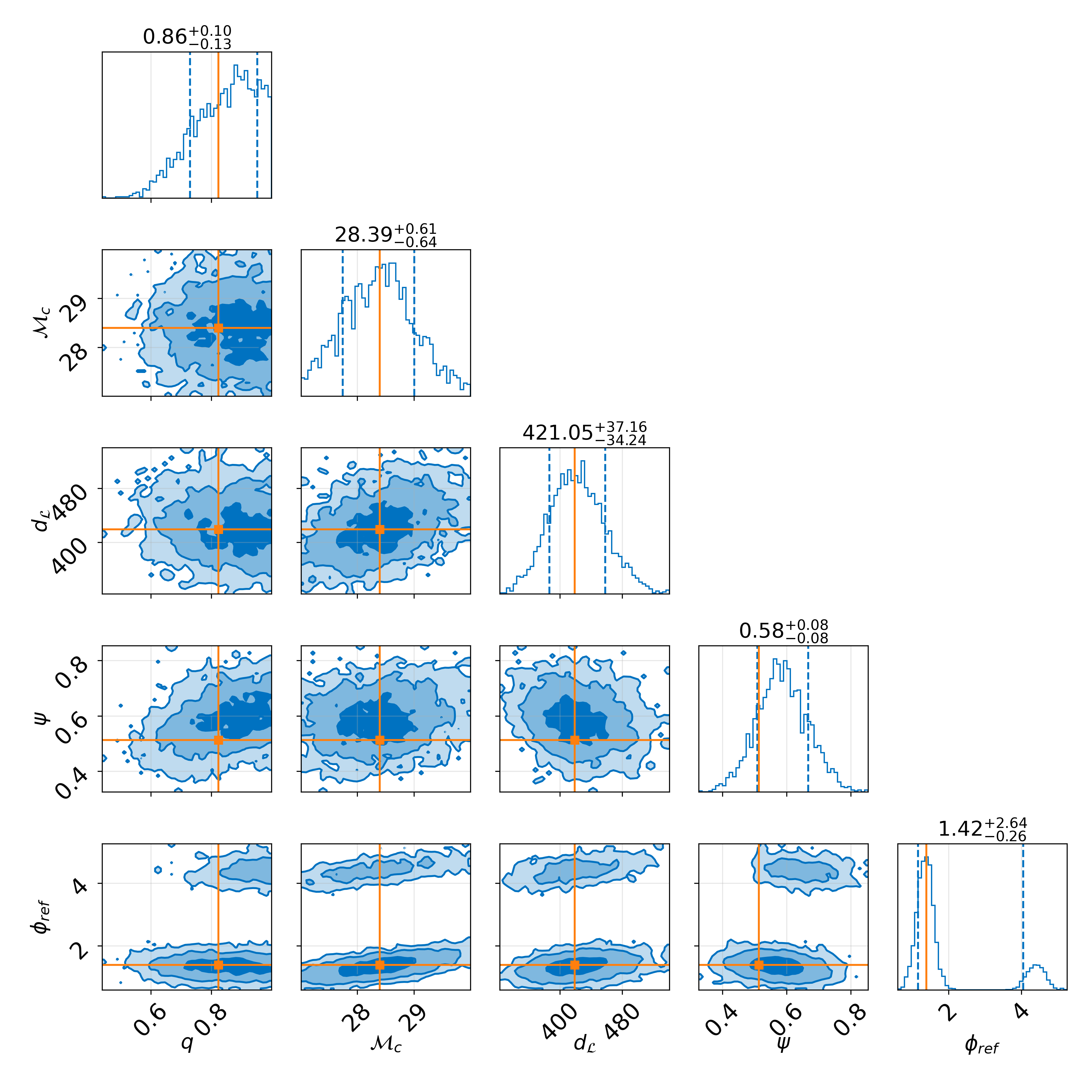}
    \caption{Injection recovery from the GPU run. This corner plot shows some of the parameters from an 11-dimensional PE run on a GPU.}
    \label{fig:gpu}
\end{figure*}
In this section, we demonstrate the PE results obtained by \emph{tdanalysis} using a GW250114-like and GW230814-like event. For the former, we truncate the signal at $-200M$ before the peak. Here, $M$ is the maximum-likelihood estimate of the total mass in the detector frame, inferred from a full IMR PE, also carried out entirely in the time domain using this pipeline. Injections are performed in both zero noise and Gaussian noise, and the injected values are chosen to be the maximum likelihood value from the IMR analysis. 
\begin{figure*}
    \centering
    \includegraphics[width=0.49\linewidth, height=0.38\linewidth]{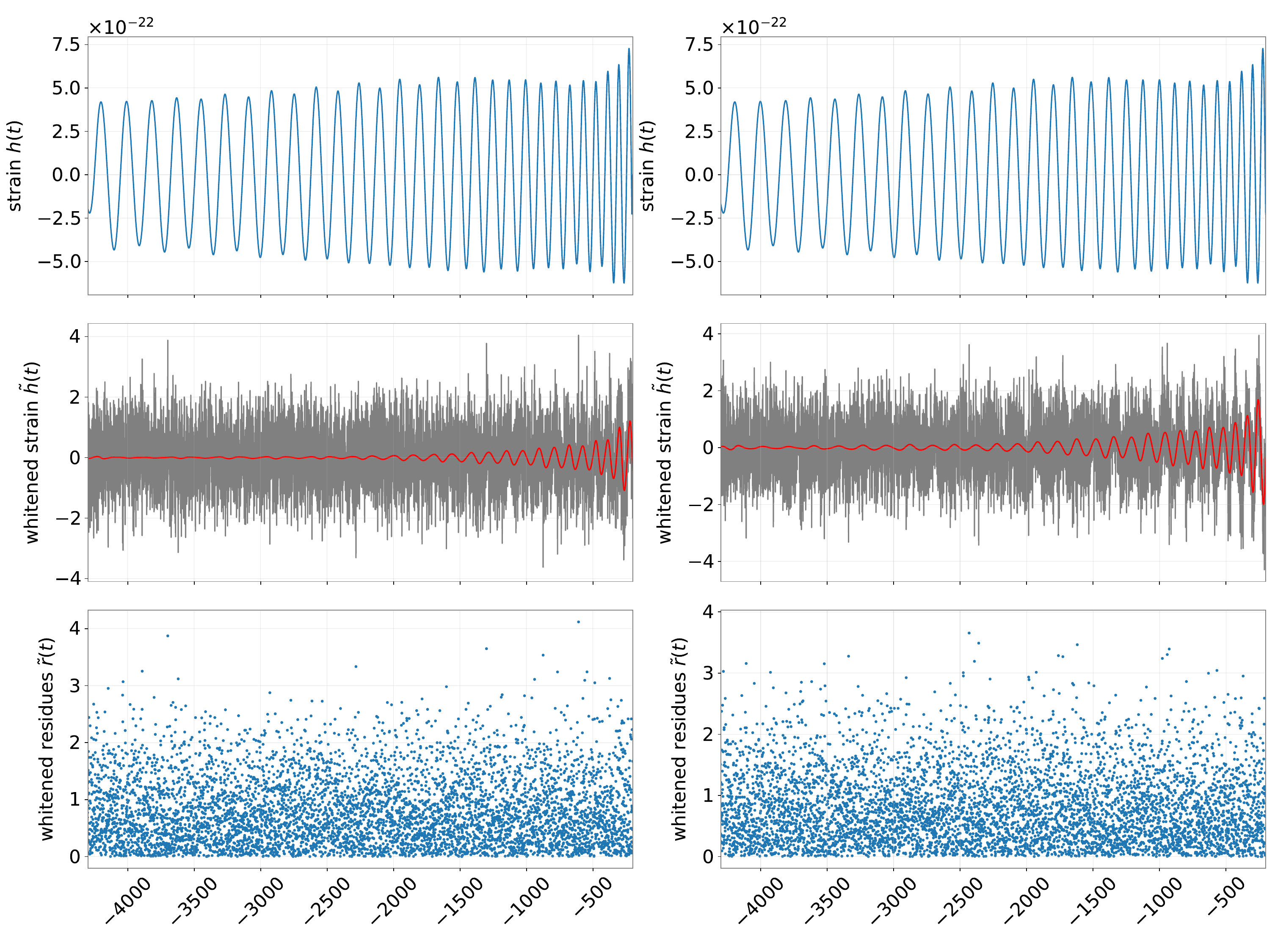}
    \includegraphics[width=0.49\linewidth]{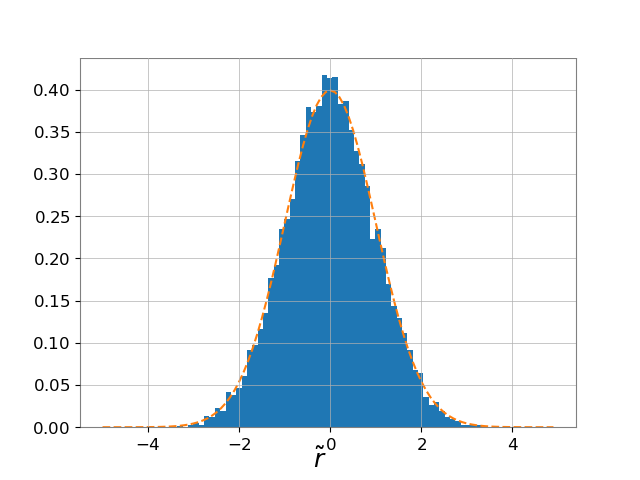}
    \caption{The recovered maximum Likelihood waveform. The top panel of the left plot shows the maximum-likelihood waveform in the Hanford detector, recovered from the GW250114 analysis of real data. The middle panel shows the whitened data along with the whitened waveform. The bottom panel shows the residues. On the right panel, the residue distribution shown in the bottom panel of the left plot is shown, along with the expected unit variance normal distribution in red.}
    \label{fig:wf}
\end{figure*}
\begin{figure*}
    \centering
    \includegraphics[width=0.24\linewidth]{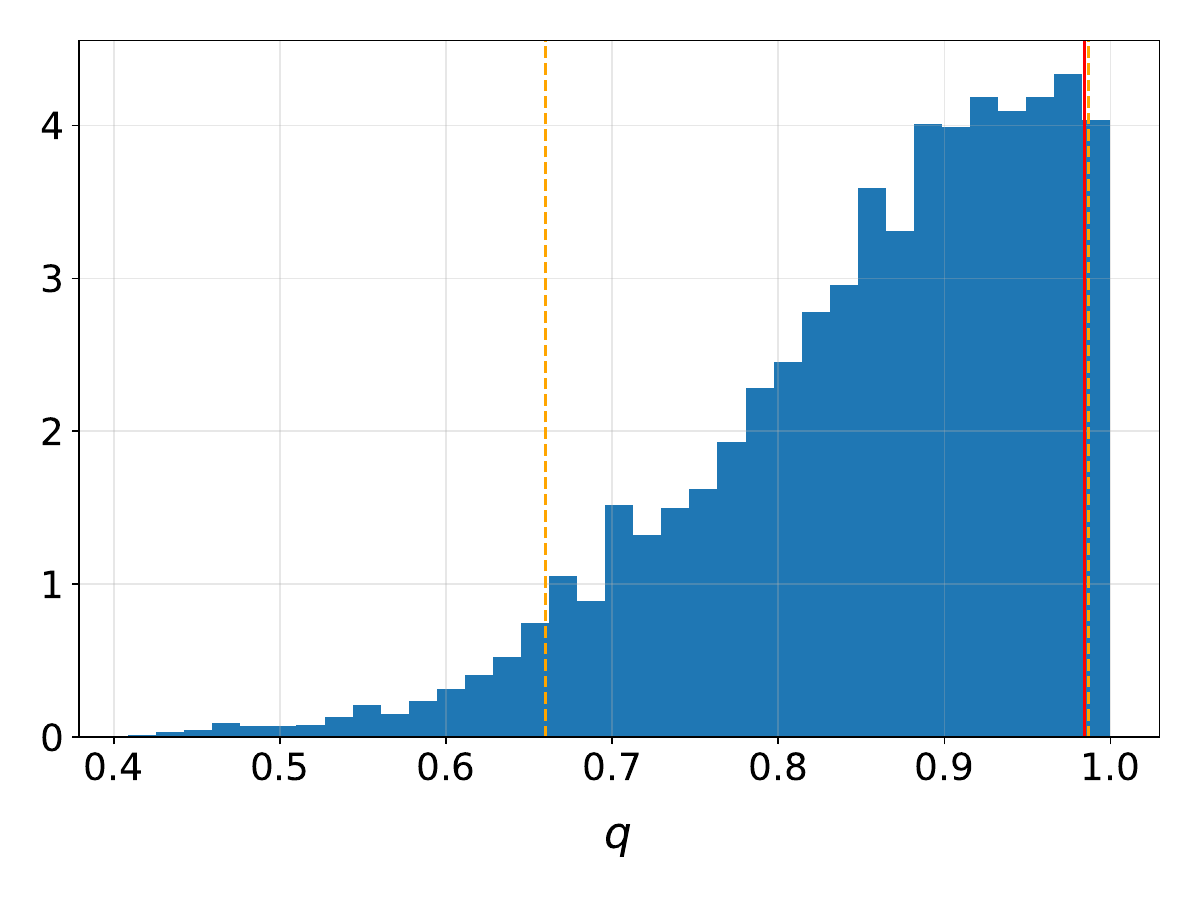}
    \includegraphics[width=0.24\linewidth]{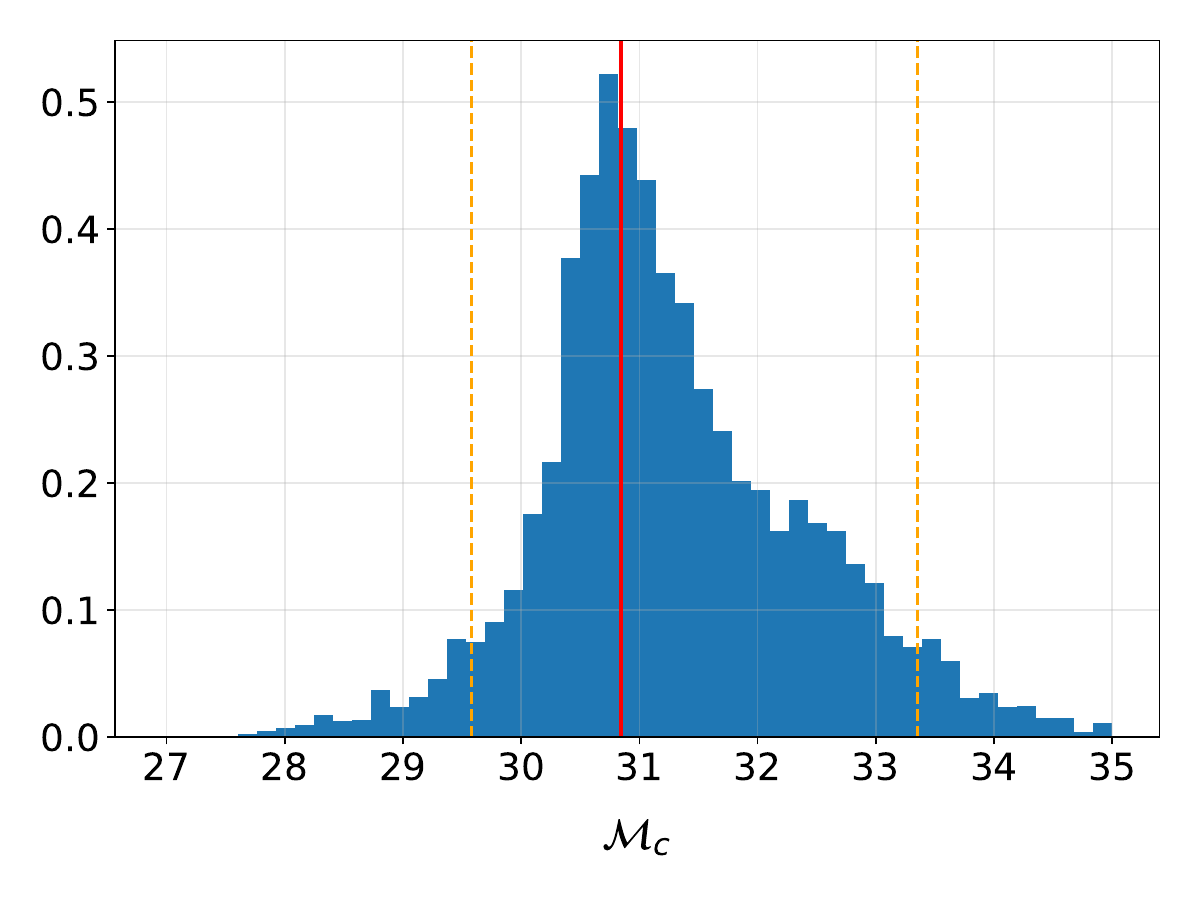}
    \includegraphics[width=0.24\linewidth]{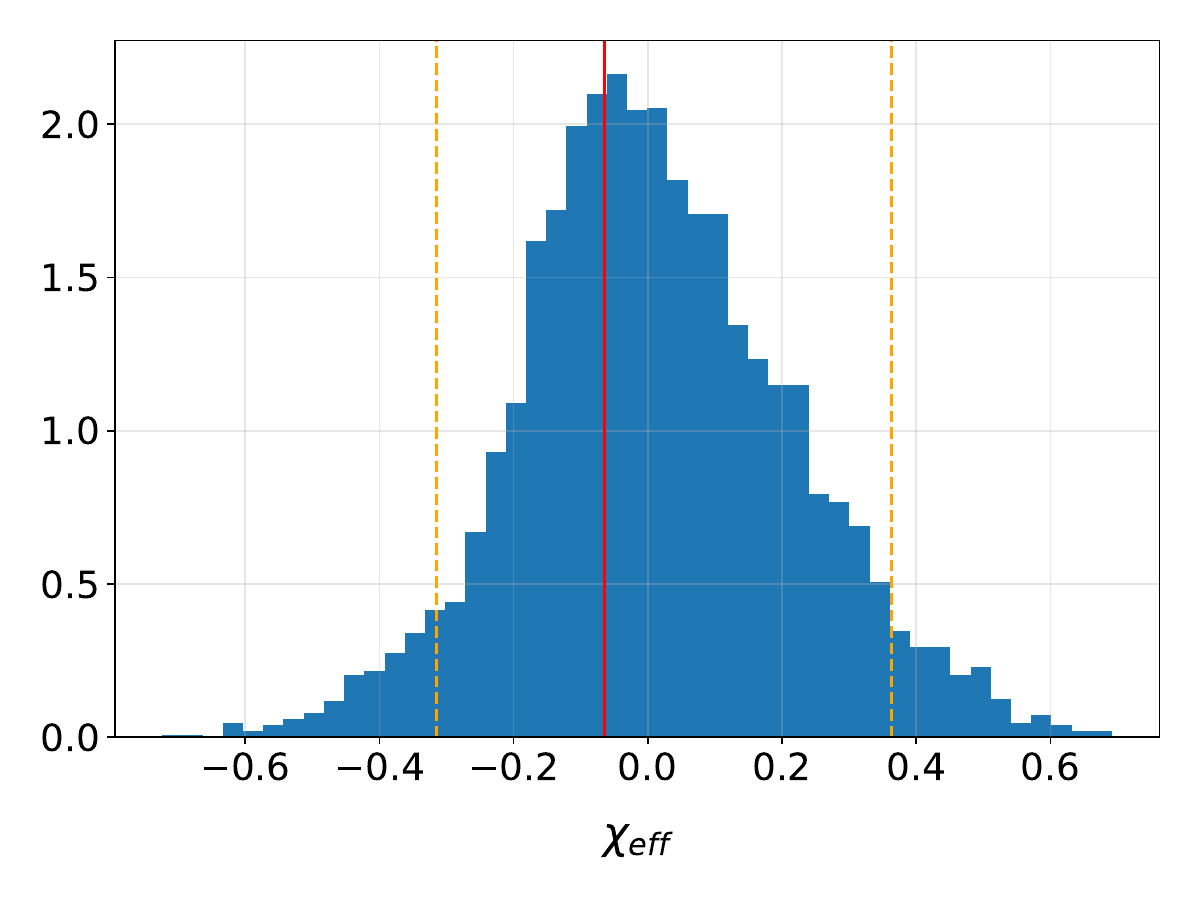}
    \includegraphics[width=0.24\linewidth]{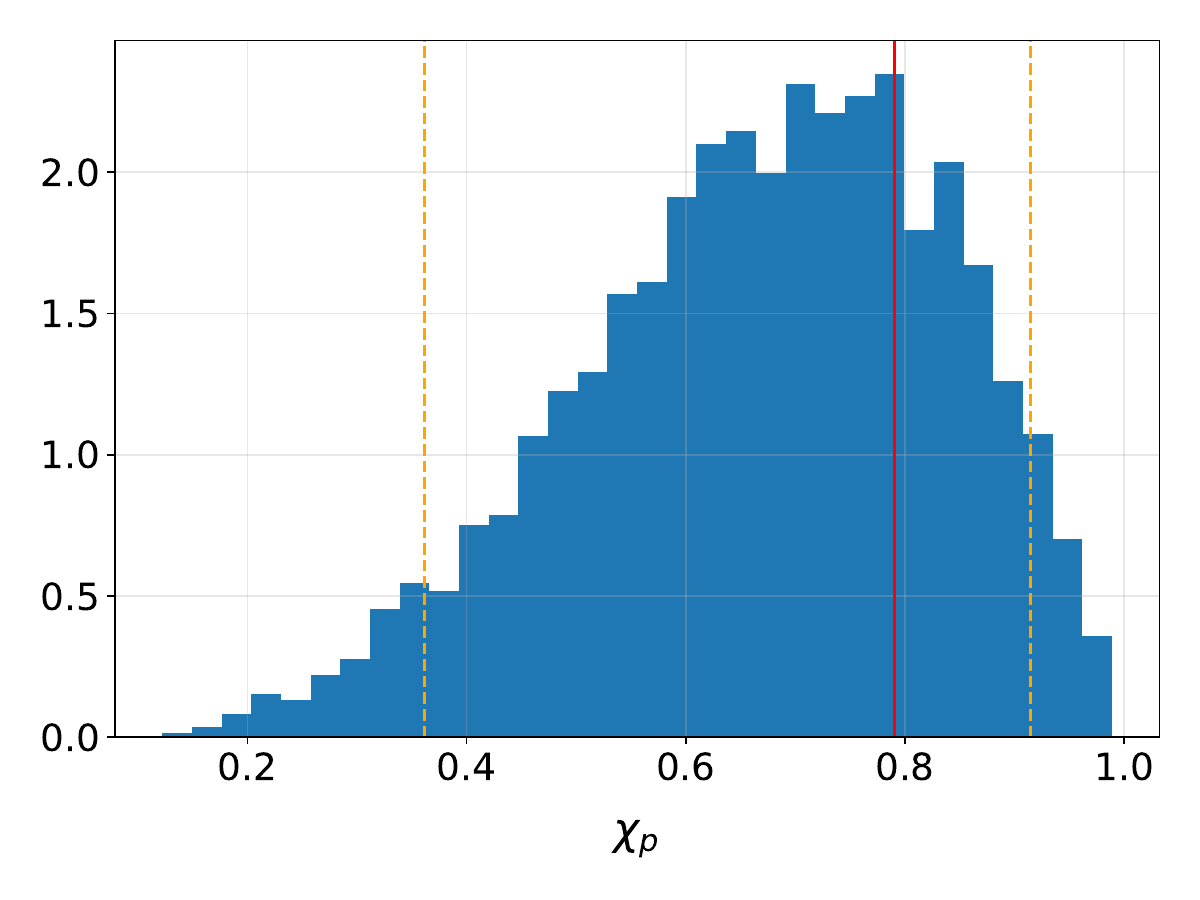}\\
 \includegraphics[width=0.24\linewidth]{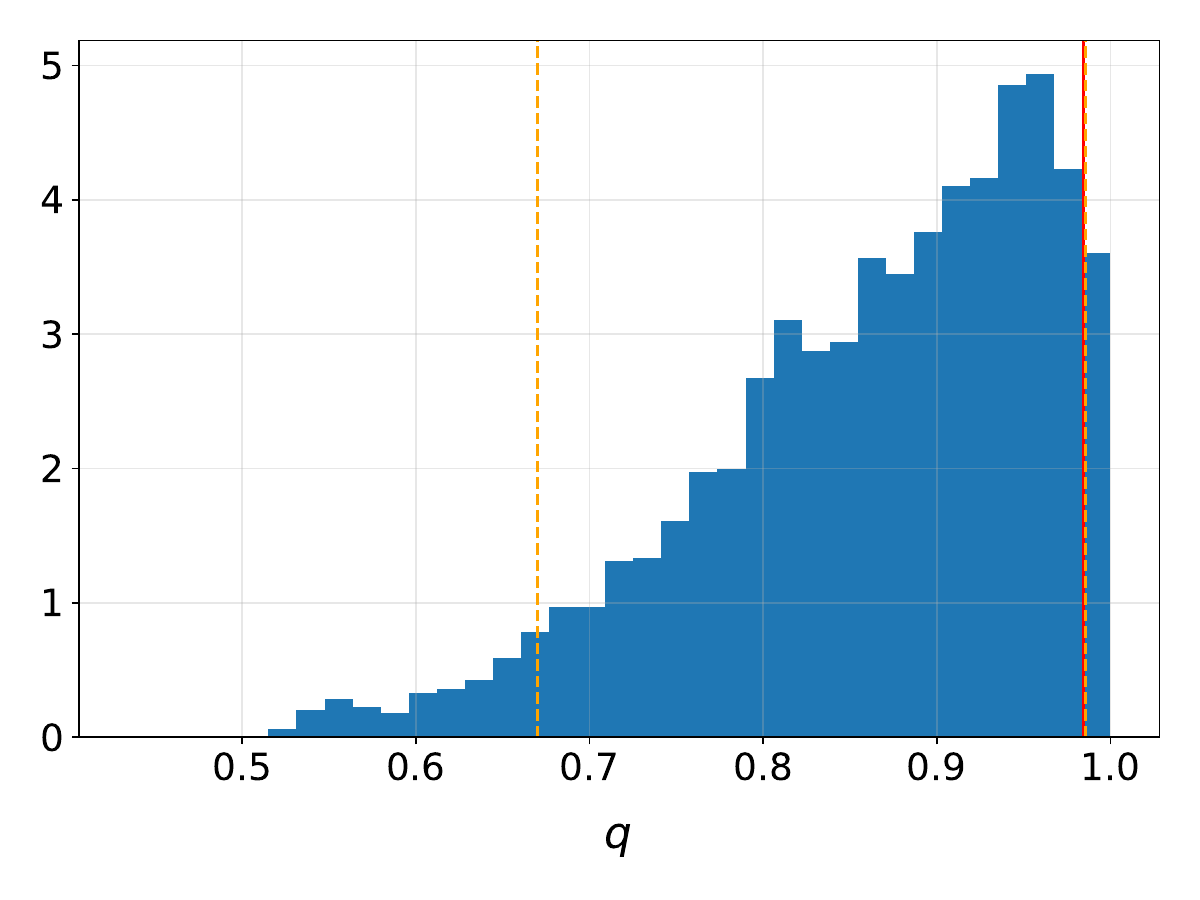}
    \includegraphics[width=0.24\linewidth]{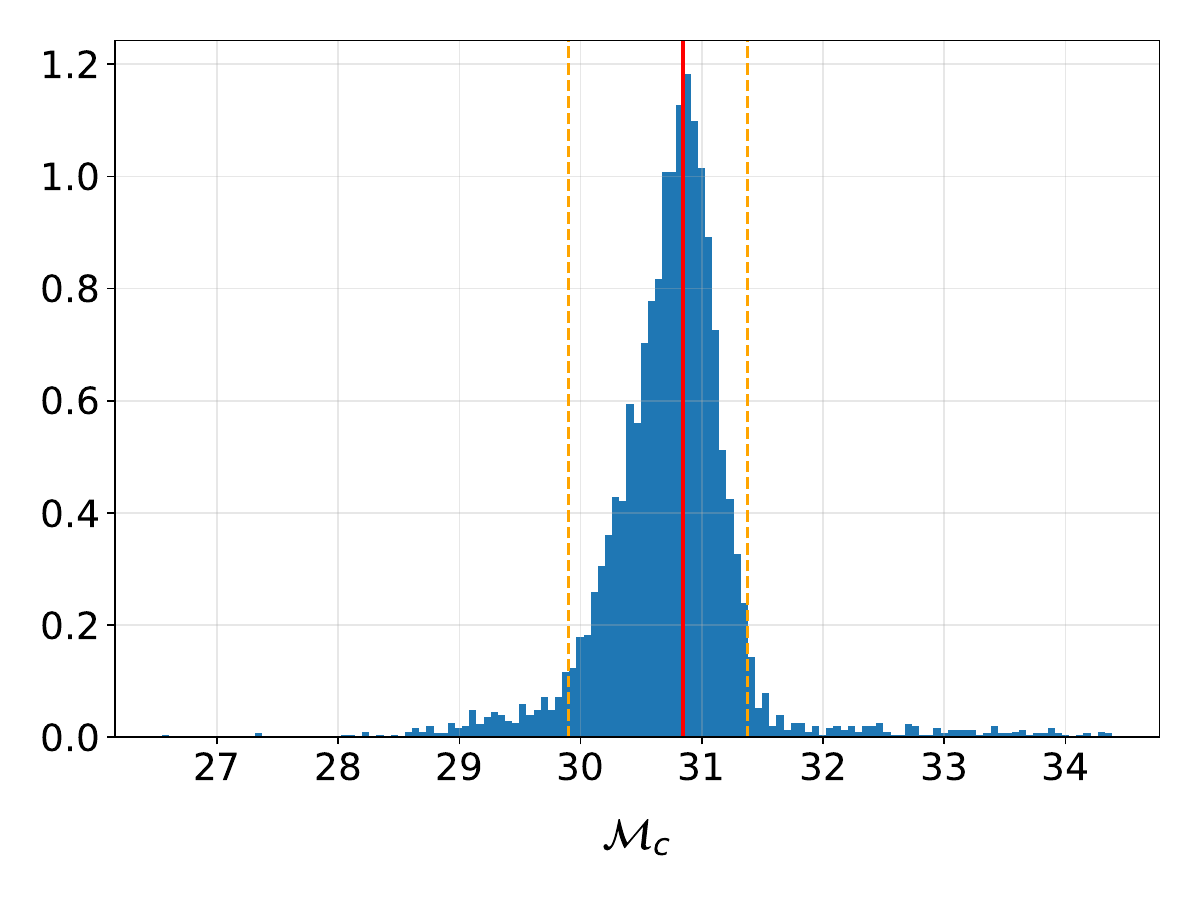}
    \includegraphics[width=0.24\linewidth]{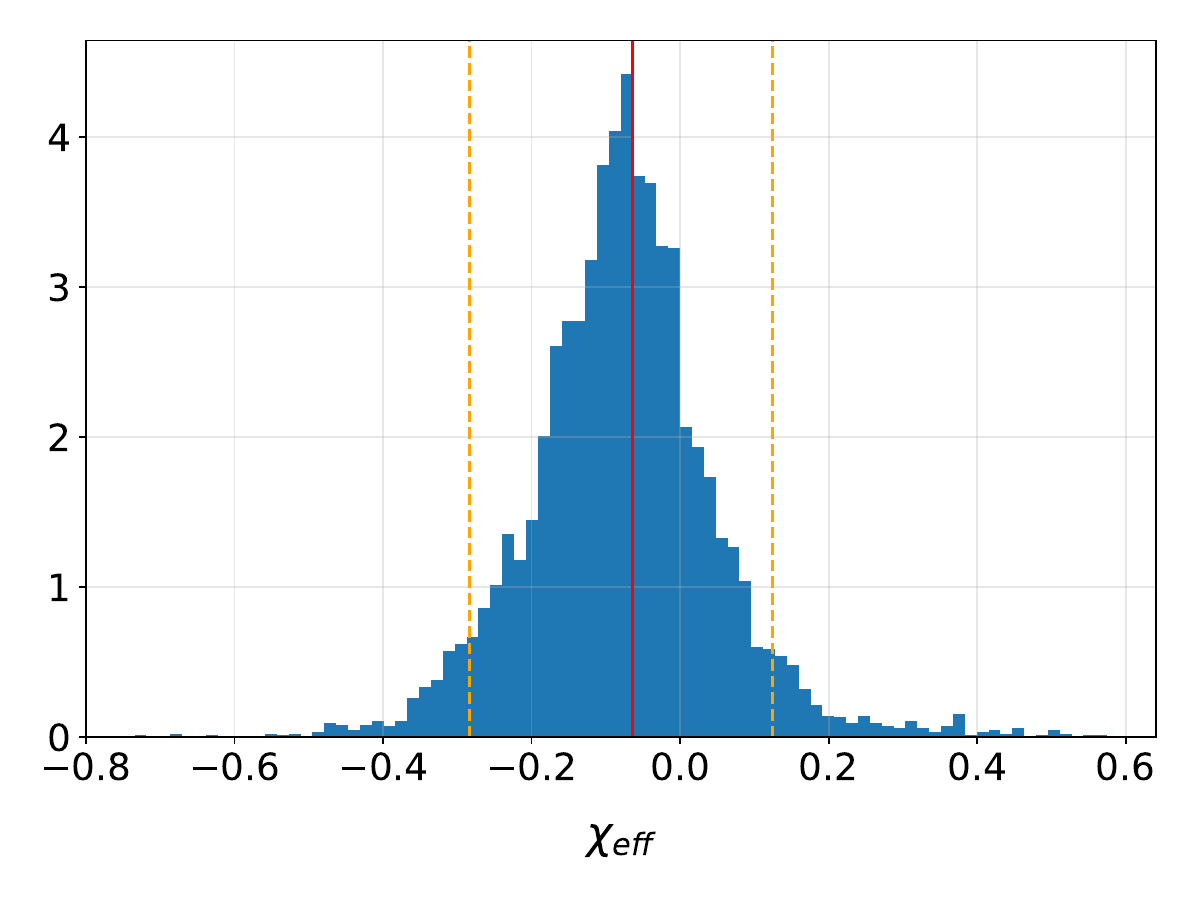}
    \includegraphics[width=0.24\linewidth]{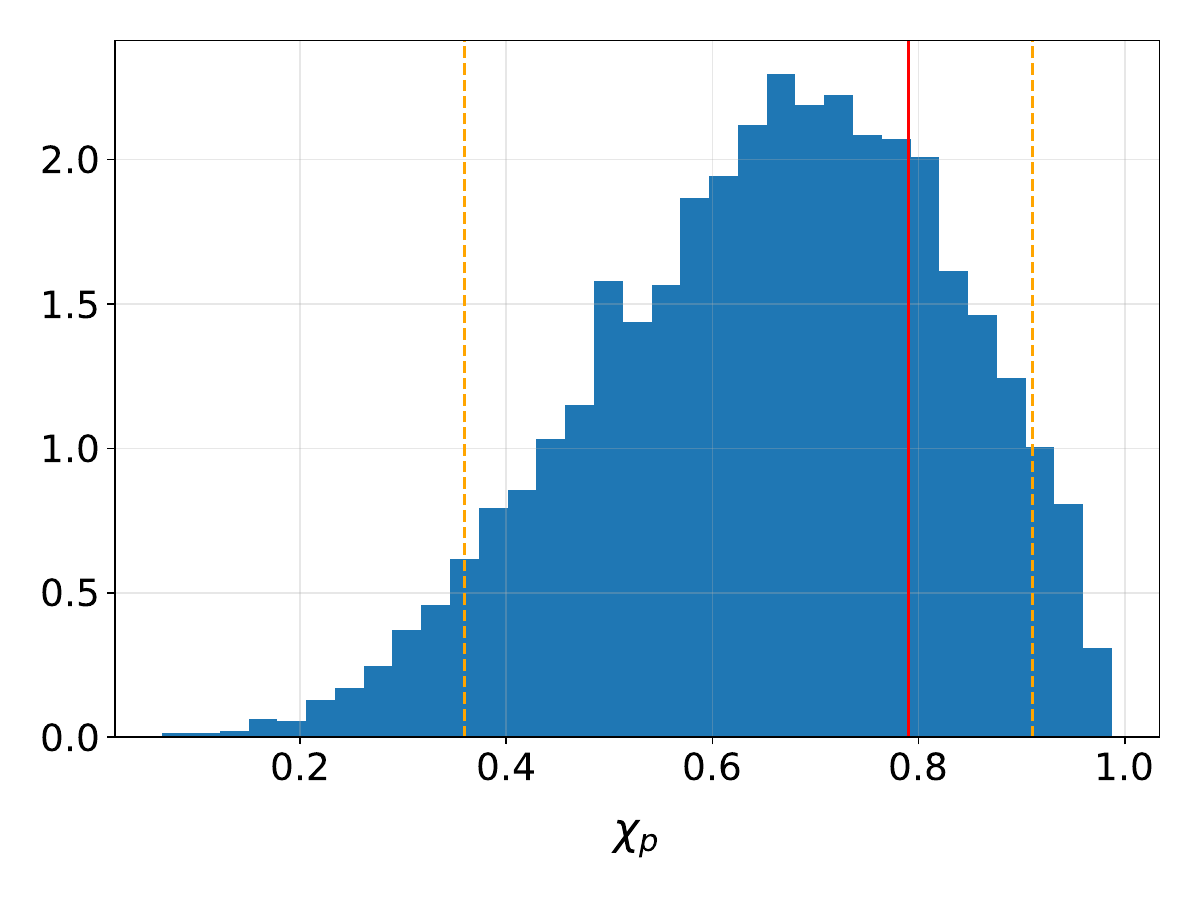}\\
     \includegraphics[width=0.24\linewidth]{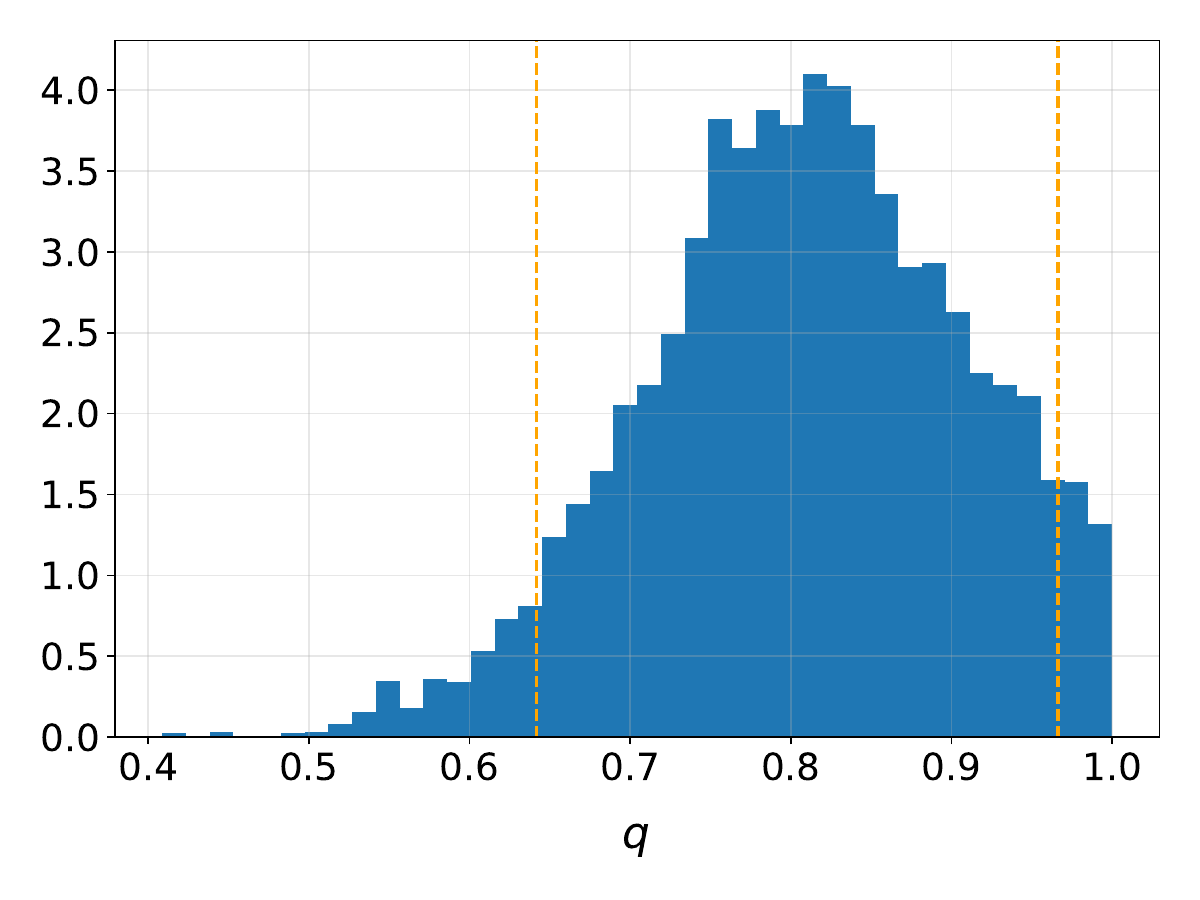}
    \includegraphics[width=0.24\linewidth]{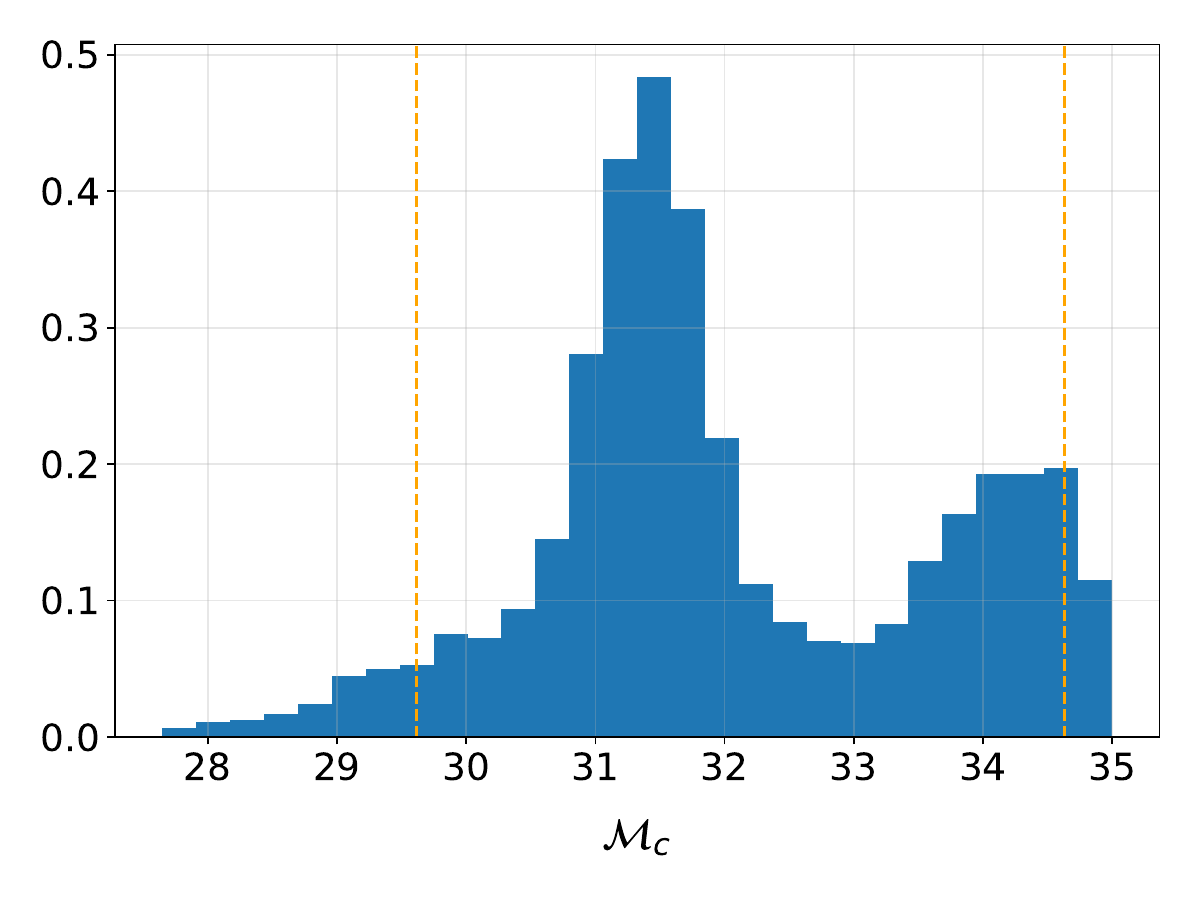}
    \includegraphics[width=0.24\linewidth]{figures/GW250114_inj200_chi_eff.pdf}
    \includegraphics[width=0.24\linewidth]{figures/GW250114_inj200_chi_p.pdf}
    \caption{Marginal distributions from GW250114 PE analysis. Here, we show the 1D marginal distributions of mass ratio (left), chirp mass (centre left), effective $\chi_{eff}$, and perpendicular spin components $\chi_p$. The top row posteriors are from the zero-noise injection; the middle row corresponds to the Gaussian synthetic noise; and the bottom row corresponds to the real data. The 90\% CI is shown as orange dotted vertical bars, and the injected value is shown as a red vertical line.}
    \label{fig:pe}
\end{figure*}
\begin{figure*}
    \centering
    \includegraphics[width=0.99\linewidth]{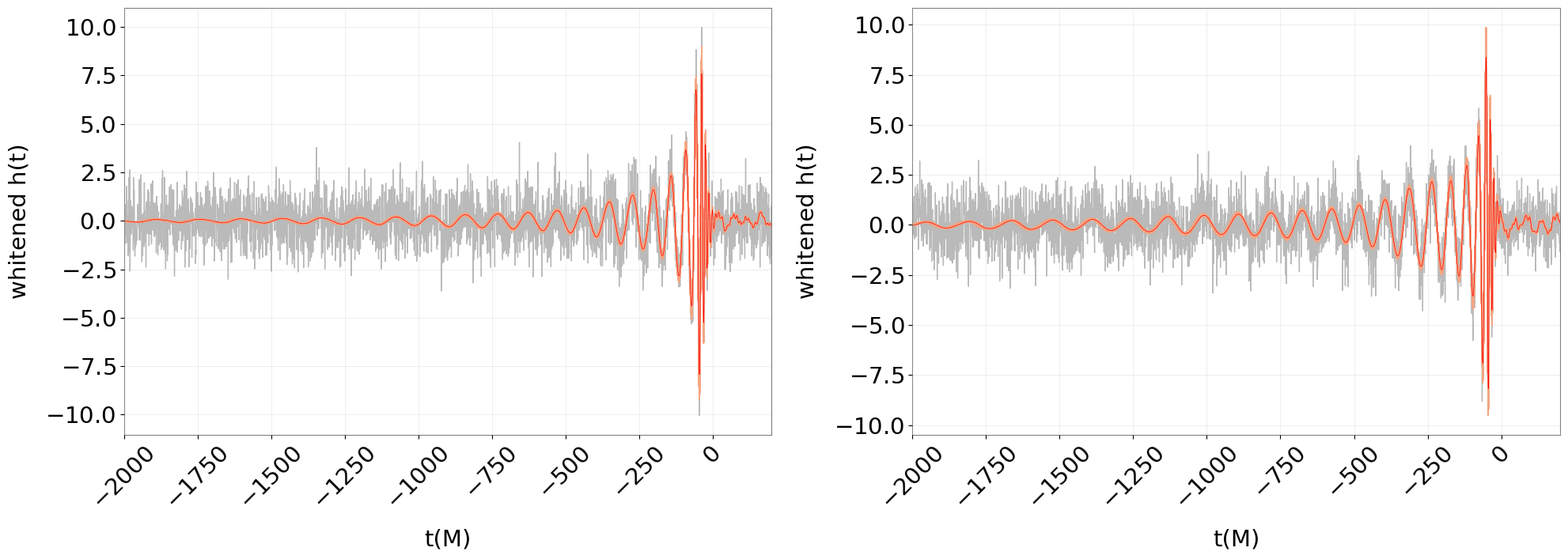}\\
    \includegraphics[width=0.99\linewidth]{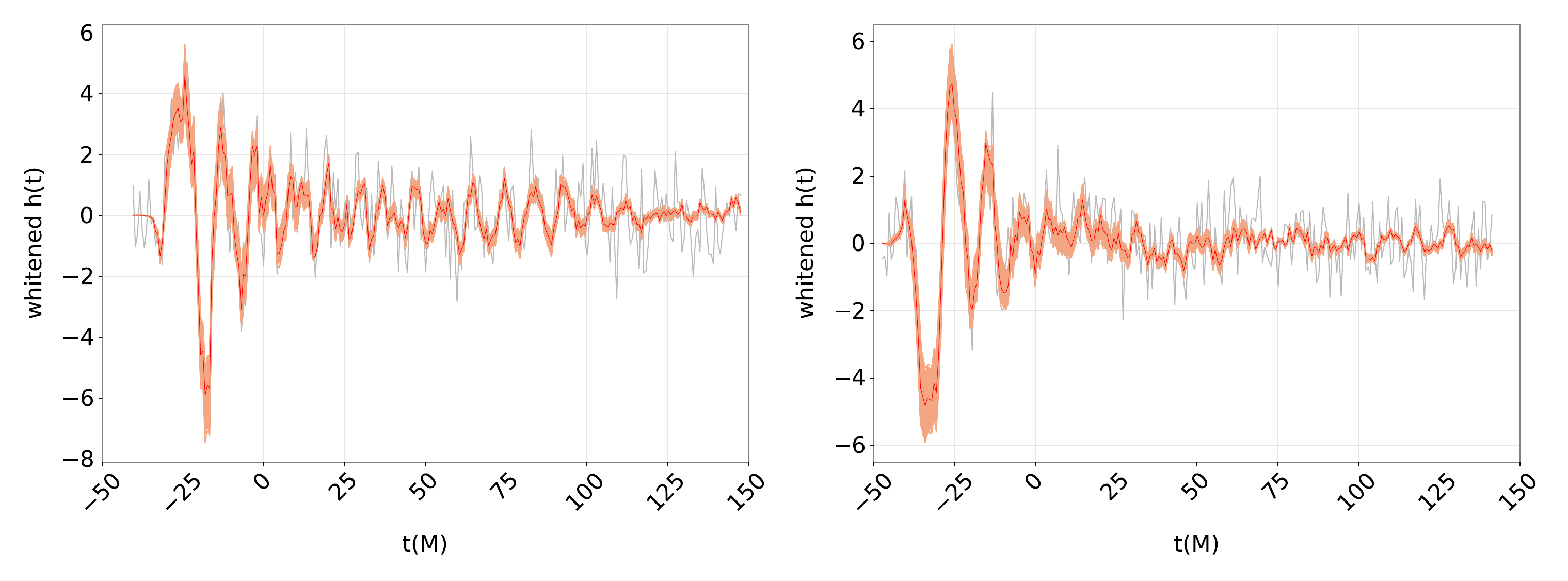}\\
    \includegraphics[width=0.99\linewidth]{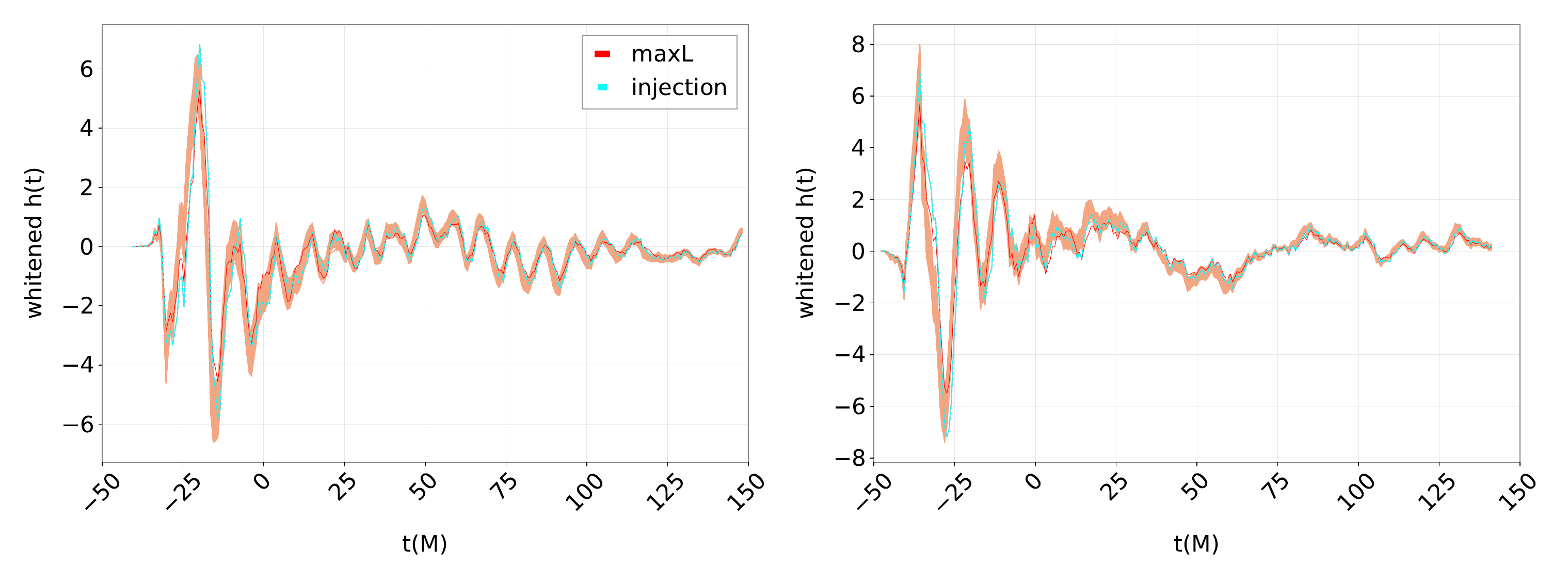}\\
    \caption{The recovered maximum Likelihood waveforms. The left panels show whitened (decorrelated and standardized) data from the Hanford detector, and the right panels from Livingston. The top panel shows the maximum-likelihood waveform recovered from the GW250114 IMR analysis in red, with the faint orange band showing the set of all posterior waveforms. The middle panel shows the same for the ringdown analysis, started 10M after the GPS peak time of the polarizations arriving at the geocentre. The bottom panel shows the corresponding plots from a zero-noise injection study, using the maximum-likelihood waveform from the full IMR analysis.}
    \label{fig:wfrecons}
\end{figure*}

\begin{figure}
    \centering
\includegraphics[width=\linewidth]{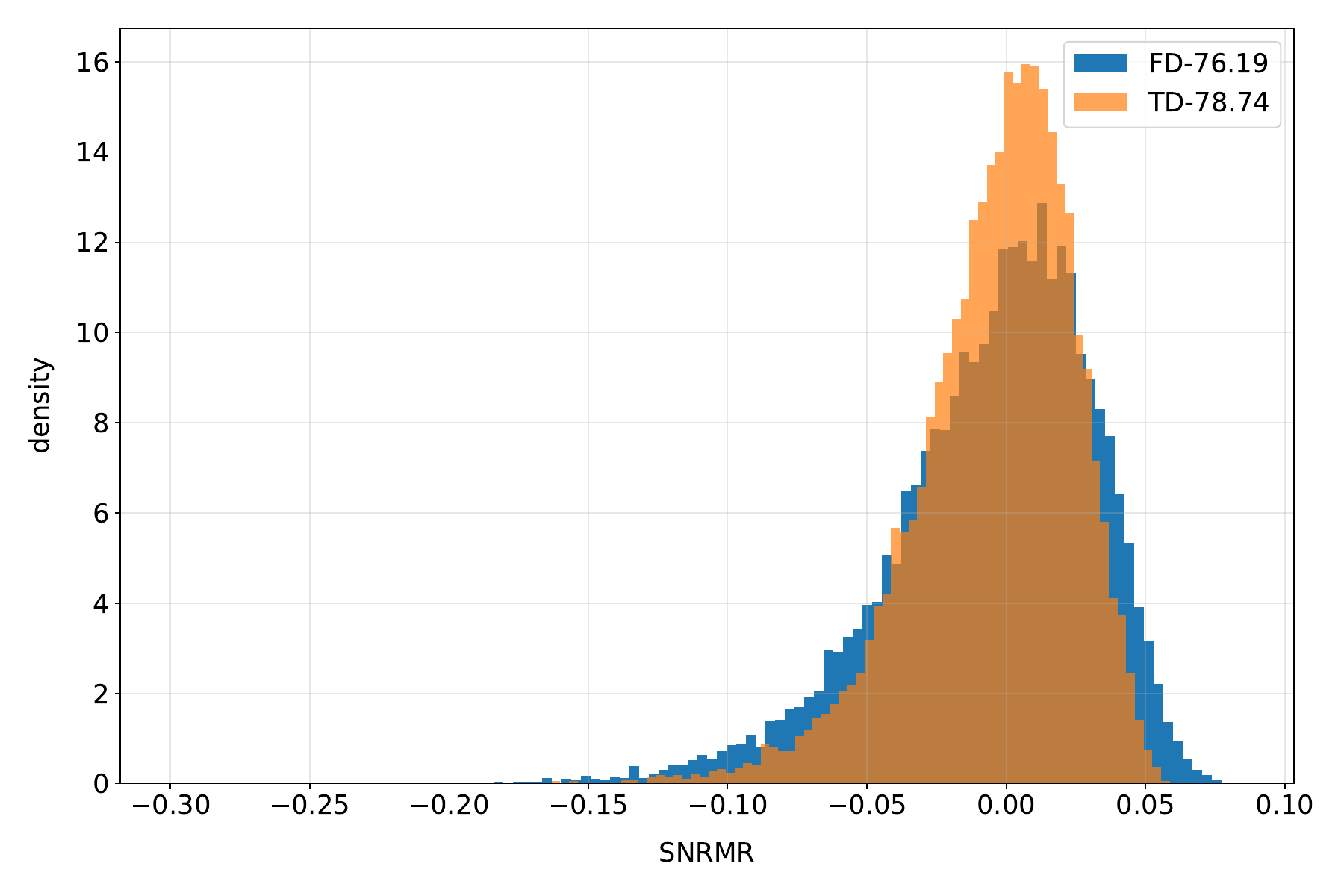}
    \caption{Matched filter SNR distributions. Here, the posterior distribution on the matched filter SNR obtained using a frequency-domain analysis (bilby) is shown, along with the corresponding distribution from this code. Note that the distributions are centered at their median values, which are mentioned in the legend.}
    \label{fig:snrs}
\end{figure}

\begin{figure*}
    \centering
    \includegraphics[width=\linewidth]{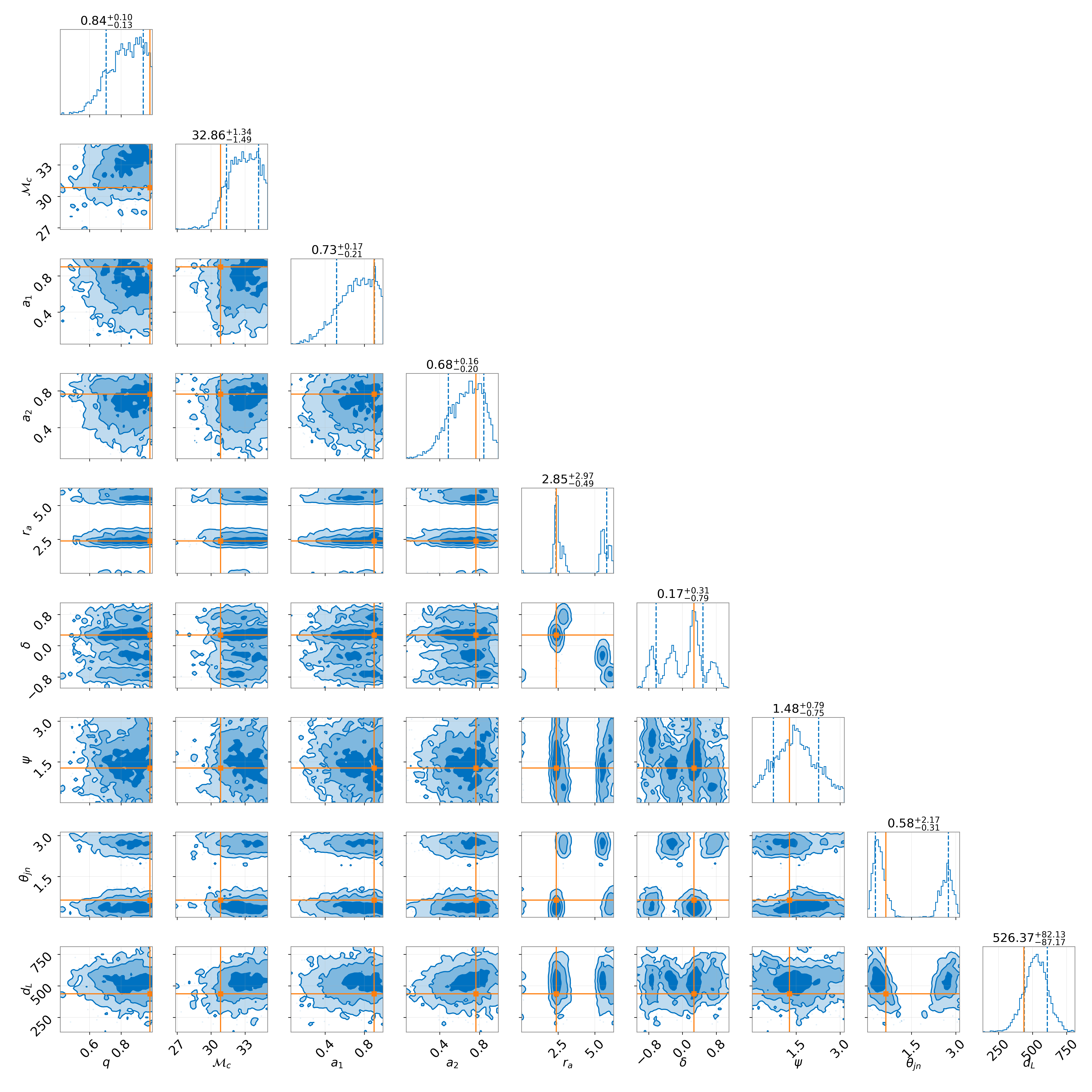}
    \caption{Recovery of parameters from a ringdown injection study. This corner plot shows the one, two, and three sigma highest-density intervals of a 2D-Gaussian distribution. The orange lines show the injected value. The rest of the 6 parameters are included in the analysis, but omitted here for brevity.}
    \label{fig:rdpost}
\end{figure*}

\subsubsection{Zero noise injection}

First, we inject a complete GW230814-like signal into the detector with zero noise and analyze it with NRSur7dq4, the waveform of choice for all the runs presented here. We run the CDO method, fixing 4 of the 11 parameters, viz. $r_a, \delta, t_{c}, \theta_{jN}$. The corner plot from some of the 11d parameters is shown in Fig.~\ref{fig:gpu}. The run took about 78.825 hours on an NVIDIA A100 GPU with 64-bit precision. The number of live points was 500, with the stopping criterion being $dlogz=0.1$.

Next, we inject GW250114 with zero noise into the Livingston detector. We terminate the signal at $-200M$ before the peak. The signal parameters are chosen to be the maximum-likelihood estimates from the IMR analysis. We run a full 15-dimensional PE using the GSCE method. Fig.~\ref{fig:pe} shows the marginal distributions of the posteriors along with the injected value. The recovery was consistent, and some of the recovered marginal distributions are shown in the first row of Fig.~\ref{fig:pe}. The SNR of the injected signal computed in the time domain was 43, while the recovered median SNR and its 90\% CI were found to be $30.51^{+0.07}_{-0.12}$. The sampling took roughly 6.5 hours with 64-bit precision, with 1000 live points and $dlogz=0.1$.

A ringdown-only analysis was also performed using a GW250114-like signal injected with `NRSur7dq4`. The ringdown portion of the injection was analyzed using the corresponding portion of the same waveform model, successfully and consistently recovering the injection. The analysis began 10M after the peak of the polarizations. The set of all whitened waveforms from the posterior distribution of samples is shown in the bottom panel of Fig.~\ref{fig:wfrecons}.

\subsubsection{Gaussian noise injection}

We repeat the analysis from the zero-noise case while also injecting noise into the detector. The noise draw, consistent with the estimated noise covariance matrix, was computed by drawing $N_{sam}$ points $\mathbf{\tilde{n}}$ from the standard normal distribution and then applying the inverse whitening filter, i.e., $\Lbf \tilde{n}$. The results of the analysis are shown in the second row of Fig.~\ref{fig:pe}.

\subsubsection{Real data}

We analyze the real data with detector noise for GW250114. We carry out three separate analyses, involving (i) the full IMR signal, (ii) the inspiral-only portion truncated at $-200M$ before merger, and (iii) the ringdown portion $10M$ after the peak of the signal.  In Fig.~\ref{fig:}. The set of all reconstructed posterior waveforms is shown in the top and middle panels. The corresponding waveforms for the inspiral-only analysis are shown in~\ref{fig:wf}

The recovered marginal distributions are shown in the third row of Fig.~\ref{fig:pe}. Some multi-modality is seen in the chirp mass-distribution, which could be caused by the signal length. The recovered maximum likelihood waveforms and the residual distribution are shown in Fig.~\ref{fig:wf}. The residual noise distribution had a Kolmogorov–Smirnov (KS) statistic value of $7.81\times 10^{-3}$ and a p-value of $0.505$ for the probability of getting a residual noise distribution at least as extreme as this from a standard normal. The Anderson–Darling test statistic, which weights the tails more heavily than the KS test, returned a value of 0.328. All these values are consistent with a standard normal distribution of residues. 

A consistent notable feature in the time domain analysis across all the analyses carried out is the recovery of higher SNRs and a higher precision thereof. Fig.~\ref{fig:snrs} shows the 1D histogram of the matched filter SNR from a frequency domain analysis using bilby in blue, and a time domain analysis result in orange. The runs are carried out with the same settings, but with different PSDs. The frequency-domain analysis uses the band-limited LVK-released PSD computed with BayesWave, and the time-domain analysis uses a median-averaged Welch PSD, inverse Fourier-transformed to construct an autocorrelation function. The histograms are centered on the respective median values, as indicated in the legend. The 90\% CI for the SNR distributions were found to be $76.187^{+0.075}_{-0.045}$ for FD, and $78.77^{+0.035}_{-0.056}$ for TD analysis, respectively. Their standard deviations were found to be $0.037$ and $0.029$, respectively. Running a Gaussianity test on the time domain whitened residues (right panel of ~\ref{fig:wf}) shows that the residues are consistent with a Gaussian, with a 
The mean of the whitened residues is $0.0021$, and the standard deviation is $\sigma_{\tilde{r}}=0.975$. This shows that the Welch PSD is overestimating the noise. Given that the SNRs scale as $\rho_{true} = \rho_{obs}/\sigma_{\tilde{r}}$, and correcting for this bias, the true suggested median SNR recovered by the time domain code is closer to $80$.

In the ringdown analysis performed using GW250114 with `NRSur7dq4`, the recovered posterior distributions were marginally consistent with those of full IMR analysis. The analysis began $10M$ after the peak of the polarizations. The set of all whitened waveforms from the posterior distribution of samples is shown in Fig.~\ref{fig:rd}. Fig.~\ref{fig:rdpost} shows a corner plot of select parameters from the full 15D posteriors. The injection values are mostly within the $2\sigma$ contours, with some lying on the edge, or outside of the 90\% CI of the corresponding marginal distribution. Given that the analysis uses heavily limited information of the signal past $10M$ after the merger, the recovery seems reasonable.

\section{Discussions and summary}
\label{sec:disc}

Due to historical reasons and limited computational capabilities, time-domain analysis with Bayesian inference was largely out of scope for signals containing more than $\sim 1024$ samples. However, with recent advances in computing infrastructure and the adoption of efficient linear-algebra algorithms, we show here that it is indeed possible to analyze gravitational waveforms entirely in the time domain.

By analyzing the cost budget for the individual operations involved in a full stochastic sampling procedure for Bayesian posteriors with a purely time-domain likelihood, we identify critical bottlenecks. In particular, we find that the direct time-domain implementation, which evaluates quadratic forms with the noise covariance matrix, is the main bottleneck, with $O(N^2)$ complexity.

We systematically consider hardware, software, and algorithmic accelerations to overcome the exorbitant computational complexity of this step in every likelihood evaluation.

On the hardware front, we find that, despite the vector capabilities of current-generation CPUs, the operation is memory-bandwidth-limited. In traditional inference, CPUs' full capabilities are not fully utilized when factored inverse covariance matrices are used for millions of likelihood evaluations via stochastic sampling. 

To circumvent this difficulty, we offload the computational step to datacenter GPUs such as the NVIDIA A100 and AMD MI100. Although these devices have much higher throughput than the CPUs, we find that the speedup does not scale with the GPU FLOPs for the same reasons as with CPUs. We show that one can achieve a speedup of about 10x, which is directly associated with the enhanced memory bandwidth on GPUs. 

Therefore, we conclude that using OpenMP on vector processors does not yield performance benefits, as the entire operation is memory-bandwidth-limited. We recommend using OMP\_NUM\_THREADS=1 and multiple walkers in the proposal step to draw new nested samples. 

Given these hardware bottlenecks, we turn to structured linear algebra and algorithmic acceleration to improve the speed of time-domain inferences. In particular, we note that Cholesky decomposition reduces the computational cost by a factor $O(\sim 2)$ but maintains the same computational complexity scaling of $O(N^2)$. We examine if triangular solves can be used to compute the whitened signals $\Lbf \mathbf{\tilde{v}} = \mathbf{v}$, as $\Lbf$ is lower triangular, and the lower\_triangular solve is highly optimized in linear algebra routines and should scale as $O(N)$. However, we find that, in most cases, many iterations are required for convergence, even with preconditioners, and there is no way to specify a good initial guess to reduce the number of iterations. 

To achieve faster algorithms, we turn to the Gohberg-Semencul theorem, which expresses the inverse of a Toeplitz matrix in terms of Toeplitz matrices. We then embed the Toeplitz matrices in a larger circulant matrix of size $2N$. As the eigen-basis of a circulant Toeplitz covariance matrix is diagonalized in the Fourier basis, we use FFTs to evaluate the quadratic form in~\eqref{eqn:ip} exactly, reducing the computational complexity to $\approx O(2N log 2N)$. Note that the circulant embedding only allows us to evaluate the quadratic form faster, and does not require the original covariance matrix to be circulant. Thus, it does not suffer from the limitations of the FD approach. We note that adapting this algorithm enables time-domain parameter estimation with roughly the same complexity scaling as the frequency domain, enabling us to use vector CPUs for acceleration. We note that using GPUs can further speed up, especially for durations greater than $O(1000)$ seconds.

We then test our pipeline on a series of injections in zero, Gaussian-simulated, and real detector noise with segmented data, demonstrating the efficiency, practicality, and accuracy of the time-domain analysis. 

Time-domain analysis has several advantages over frequency-domain analysis. The most important of all is the ability to analyze time-localized segments of the signal, sharply truncated at either end, without the need to apply window/ tapering functions.

Further, time-domain analysis has the distinctive advantage of potentially generalizing to non-stationary signals, which we explore in a following work.


\begin{acknowledgments}

The author thanks B.S. Sathyaprakash for continuous support and encouragement.

This research is supported by National Science Foundation grants Nos.\, PHY-2308886, and No. PHY-2309064, and PHY-250294. The early version of this pipeline was developed by the author at IUCAA, Pune (see e.g., its application in \cite{Prasad:2023bwa}), as a PhD student, with support from the CSIR Shyama Prasad Mukherjee Fellowship.

Further development of this pipeline was carried out by the author as a Post-doctoral Fellow at ICTS-TIFR, supported by the Department of Atomic Energy, Government of India, under project no. RTI4001.

This material is based upon work supported by NSF's LIGO Laboratory, which is a major facility fully funded by the National Science Foundation

Facilities: EGO:Virgo, GEO600, Kamioka:KAGRA, LIGO

Software: Plots were prepared with Matplotlib \cite{Hunter2007Matplotlib}, waveforms were generated through LALSuite \cite{LVK2018LALSuite, Wette2020SWIGLAL}.  NumPy \cite{harris2020array}, SciPy \cite{Virtanen2020SciPy} were used for data processing in generating the figures and quantities in the manuscript. Bilby \cite{Ashton:2018jfp, RomeroShaw2020BilbyValidation, Smith2020pBilby} and Dynesty \cite{Speagle:2019ivv, Koposov2025dynesty_v3_0_0} for stochastic sampling with Bayesian Inference. The time domain code used was developed by the author, beginning with his PhD thesis. \emph{tdanalysis} is expected to be released to the public domain soon, but is currently available for use upon request.

Computing: The authors acknowledge the computational resources provided by the NSF Access grant PHY-250294, and partly by the LIGO Laboratory’s CIT cluster, which is supported by National Science Foundation Grants PHY-0757058 and PHY-0823459, and the Gwave cluster at Pennsylvania State University (supported by NSF grants OAC-2346596, OAC-2201445, OAC-2103662, OAC-2018299, and PHY-2110594).

This work primarily used Bridges-2 at the Pittsburgh Supercomputing Center through allocation PHY-250294 from the Advanced Cyberinfrastructure Coordination Ecosystem: Services \& Support (ACCESS) program, which is supported by the National Science Foundation grants No. 2138259, 2138286, 2138307, 2137603, and 2138296. This research also used the Delta advanced computing and data resource, which is supported by the National Science Foundation (award OAC 2005572) and the State of Illinois. Delta is a joint effort of the University of Illinois Urbana-Champaign and its National Center for Supercomputing Applications.

\end{acknowledgments}

\bibliographystyle{apsrev4-2}
\bibliography{references_norm}

\end{document}